\documentclass[twocolumn,showpacs,amsmath,amssymb,aps]{revtex4}
\usepackage{graphicx}
\usepackage{dcolumn}
\usepackage{epsfig}
\usepackage{bm}
\usepackage{gensymb}
\usepackage{textcomp} 
\usepackage{hyperref}
\usepackage{color}

\begin{document}

\title{Low-energy excitations and quasielastic contribution to electron-nucleus and neutrino-nucleus scattering in the
continuum random phase approximation}

\author{V.~Pandey\footnote{Vishvas.Pandey@UGent.be}, N.~Jachowicz\footnote{Natalie.Jachowicz@UGent.be}, T.~Van Cuyck, 
J.~Ryckebusch, M.~Martini}

\affiliation{Department of Physics and Astronomy,\\ Ghent University, \\Proeftuinstraat 86, \\ B-9000 Gent, Belgium.\\}


\begin{abstract}

We present a detailed study of a continuum random phase approximation approach to quasielastic electron-nucleus
and  neutrino-nucleus scattering. We compare the ($e,e'$) cross-section 
predictions with electron scattering data for the nuclear targets $^{12}$C, $^{16}$O, and $^{40}$Ca, 
in the kinematic region where quasielastic scattering is expected to dominate. We examine the longitudinal and 
transverse contributions to $^{12}$C($e,e'$) and compare them with the available data. We find an
overall satisfactory description of the ($e,e'$) data. 
Further, we study the $^{12}$C($\nu_{\mu},\mu^{-}$) cross sections relevant for accelerator-based neutrino-oscillation experiments. 
We pay special attention to low-energy excitations which can account for non-negligible contributions in measurements, 
and require a beyond-Fermi-gas formalism. 

\end{abstract}



\maketitle


\section{Introduction}

The quest for a completion of our knowledge of neutrino-oscillation parameters has made tremendous progress in 
recent years. Still, neutrino-oscillation experiments face a number of challenges. Major issues are the identification 
of the basic processes contributing to the neutrino-nucleus signal in a detector and the reduction of the systematic 
uncertainties. A thorough understanding of the complexity of the nuclear environment and its electroweak response at low 
and intermediate energies is required. Charged-current quasielastic (CCQE) processes account for a large share of the detected 
signals in many experiments. Although several cross section measurements have been performed 
~\cite{miniboone:ccqenu, miniboone:ccqeantinu, miniboone:nunc, minerva:nu, minerva:antinu, t2k:qenu, t2k:qenu2}, uncertainties 
connected to the electroweak responses persist~\cite{rev:Morfin,rev:Nieves}.

Despite substantial progress in the understanding of the different processes involved in the signal of neutrino-oscillation 
experiments, the simulation codes are primarily based on a Fermi-gas description of the nucleus. Relativistic Fermi-gas 
(RFG) based models are employed in Monte Carlo event generators. The RFG model provides a basic picture of the nucleus as a system 
of quasifree nucleons and takes into account the Fermi motion and Pauli blocking effects. The analysis of electron-scattering 
data suggests that at momentum transfers $q \approx$ 500 MeV/c, the RFG model describes the general behavior of the quasielastic 
(QE) cross section sufficiently accurately, but its description becomes poor for smaller momentum transfers, where nuclear effects 
are more prominent. Since the neutrino flux in the oscillation experiments 
is distributed over energies from very low to a few GeV, the cross section picks up contributions from all energies. The low excitation-energy 
cross sections do not receive proper attention in an RFG description. Furthermore, even at higher incoming 
neutrino energy, the contributions stemming from low transferred energies are not negligible. At low energy transfers, the  
nuclear structure certainly needs a beyond RFG description. Several studies emphasizing the low energy 
excitation in the framework of neutrino-nuclear interactions~\cite{Kolbe:1995, Volpe:2000, 
Co:2005, Martini:2007, Samana:2011} have been 
performed. Those studies, however, have not been explicitly extended to explore the kinematics of MiniBooNE~\cite{MiniBooNE}, T2K~\cite{T2K}, 
and other similar experiments. 

In this paper, we present a continuum random phase approximation (CRPA) approach for the description of QE electroweak scattering 
off the nucleus, 
crucial for accelerator-based neutrino-oscillation experiments. We pay special attention to low-energy nuclear excitations. 
In this context, the availability of a large amount of high-precision electron-nucleus scattering data is of the
utmost importance, as it allows one to test the reliability of the reaction model.

Several models have been developed to study electron-nucleus scattering and further generalized to describe
neutrino-nucleus cross sections~\cite{Alberico:1982, Martini:2009, Martini:2013, Nieves:1997, Nieves:2004, Nieves:2005, Benhar:1994, Benhar:2005, 
Benhar:2007, Ankowski:2014, Amaro:2005, Amaro:2007, Mosel:2009, Butkevich:2005, Butkevich:2007, Meucci:2003, Meucci:2004, Co:1984}. An extensive test against the inclusive 
quasielastic electron scattering is performed within an RFG and plane-wave impulse approximation approach in Ref.~\cite{Butkevich:2005}, while a 
spectral function based approach is assessed in Ref.~\cite{Ankowski:2014}. The  model we adopt takes a Hartree-Fock (HF) description of nuclear 
dynamics as a starting point and additionally implements long-range correlations 
through a CRPA framework with an effective Skyrme nucleon-nucleon two-body interaction. We solve the CRPA equations by a Green's function approach. 
Thereby, the polarization propagator is approximated by an iteration of its first-order contribution. In this way, the formalism implements the
description of one-particle one-hole excitations out of the correlated nuclear ground state. To improve our description of the kinematics of the
interaction at intermediate energies, we implemented an effective relativistic approach proposed in Refs.~\cite{Donnelly:1998,Amaro:2005,Amaro:2007}. 

The article is organized as follows. In Sec.~\ref{formalism}, we outline the details of the QE electron and neutrino-nucleus cross-section  
formalism. We describe the CRPA framework for calculating nuclear responses. Sec.~\ref{results}, is divided into two parts: In 
Sec.~\ref{results_electron}, we present numerical results of electron-scattering cross sections (on a variety of nuclear targets), and
responses (on $^{12}$C) and compare them with the available data. In Sec.~\ref{results_neutrino}, we discuss neutrino-scattering 
results in the context of accelerator-based neutrino-oscillation experiments. We pay special attention to low-energy neutrino-induced 
nuclear excitations. Conclusions can be found in Sec.~\ref{conclusions}.


\section{Formalism}\label{formalism}

In this section, we describe our CRPA-based approach for the calculation of the nuclear response for inclusive electron and
neutrino-nucleus scattering in the QE region. This approach was successful in describing  exclusive photo-induced and  
electron-induced QE processes~\cite{Jan:1988,Jan:1989}, and  inclusive neutrino scattering at supernova energies~\cite{Natalie:nc1999, 
Natalie:cc2002, Natalie:nc2004, Natalie:sn2006, Natalie:proc2009, Natalie:nuintproc2012}. We have also used this approach to calculate the 
inclusive CCQE antineutrino-nucleus scattering cross sections at intermediate energies~\cite{Vishvas:ccqeantinu2014}. 
Here, we are using an updated version of the same formalism.
\begin{figure}
\includegraphics[width=0.49\columnwidth]{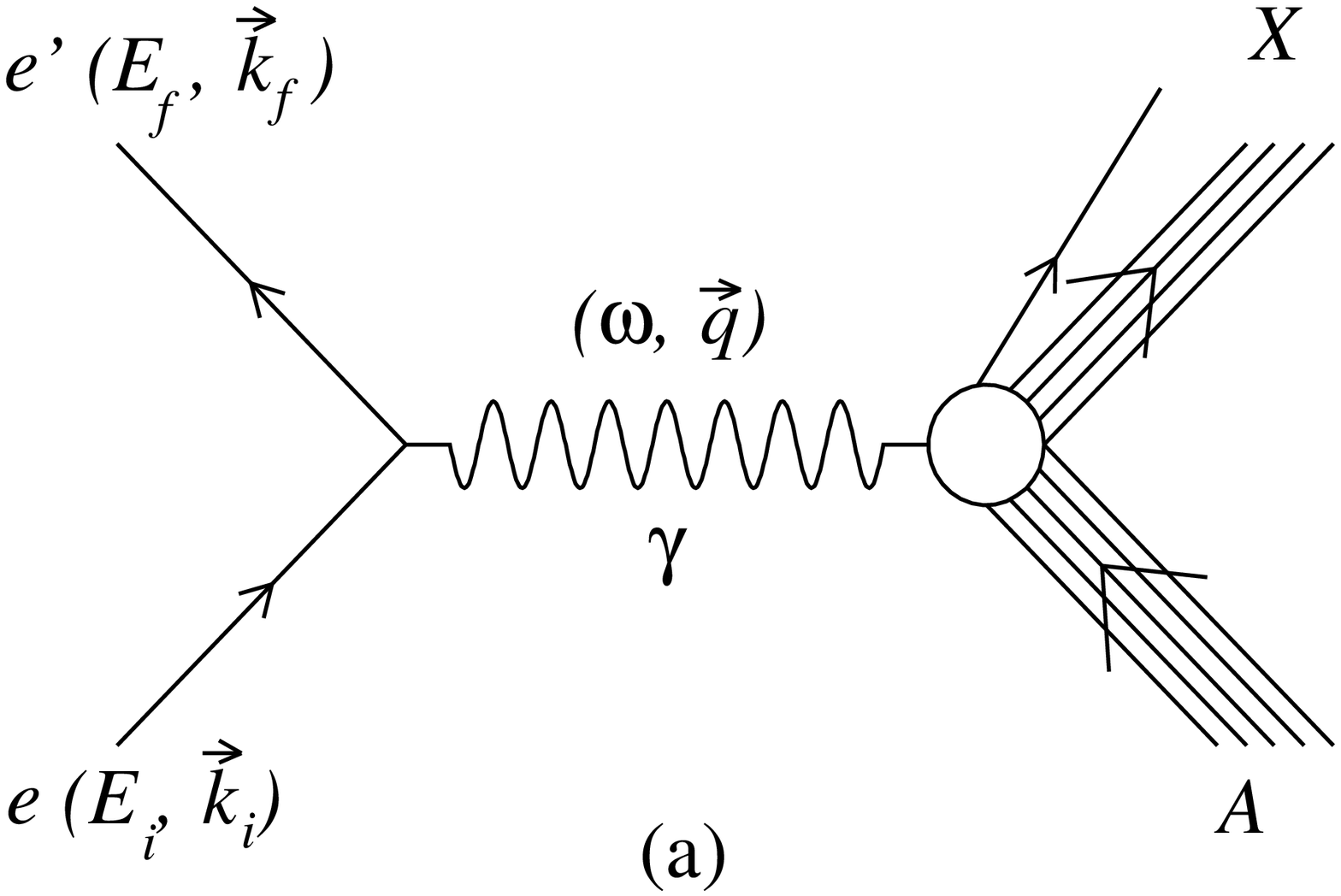}
\includegraphics[width=0.49\columnwidth]{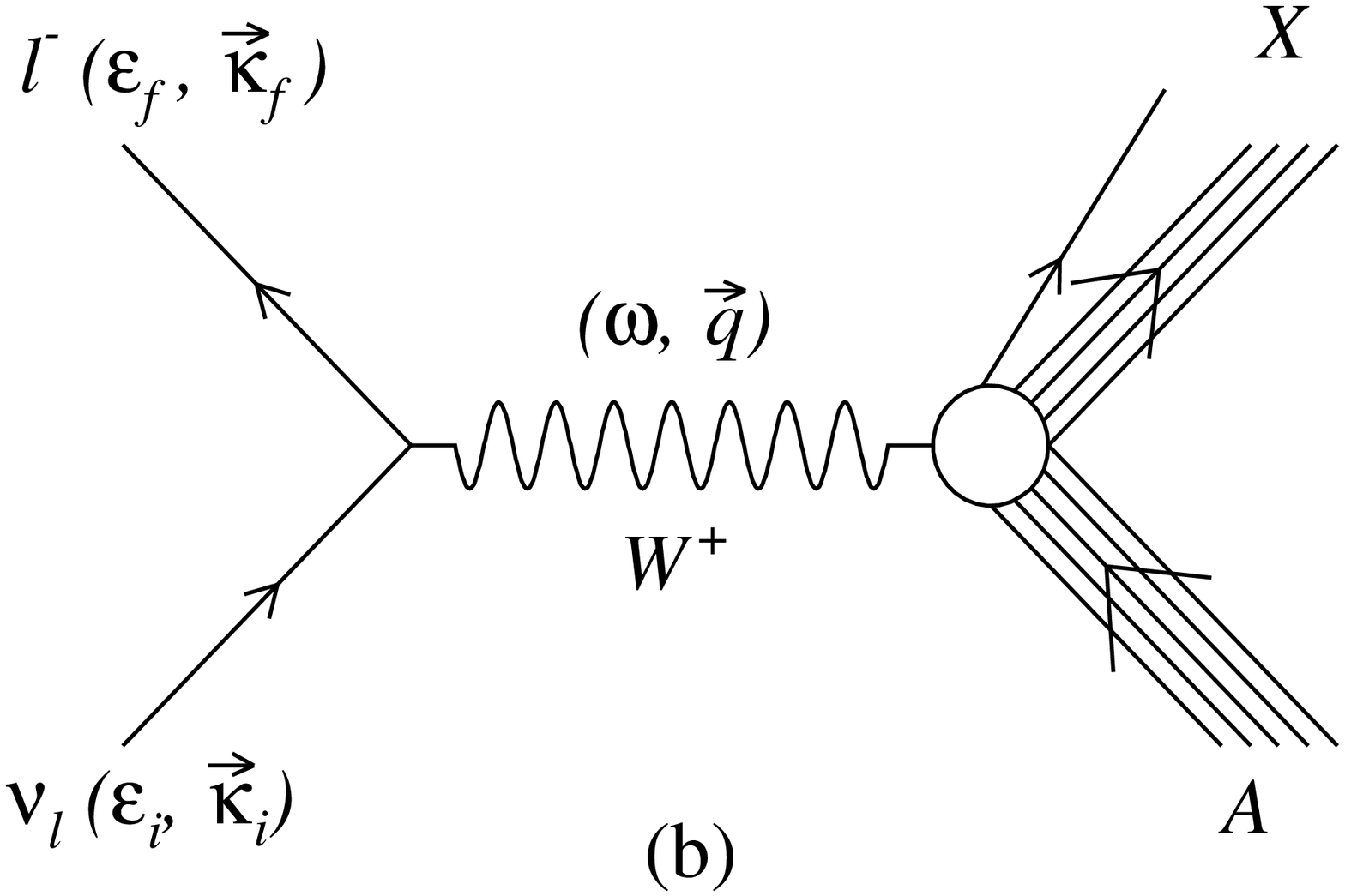}
\caption{Inclusive processes considered in this paper: (a) QE electron-nucleus and (b) CCQE neutrino-nucleus
($l = e, \mu, \tau$), where $X$ is the undetected hadronic final state.}
\label{feymann_diagrams}
\end{figure}

We consider QE electron and CCQE neutrino scattering off a nucleus under conditions where the details of the final hadron state 
remain unobserved. As shown in Fig.~\ref{feymann_diagrams}, an incident electron (neutrino) with four-momentum $E_i, 
\vec{k}_i$ ($\varepsilon_i, \vec{\kappa}_i$) scatters off a nucleus via the exchange of a photon ($W$-boson) and only 
the outgoing charged lepton with four-momentum $E_f, \vec{k}_f$ ($\varepsilon_f, \vec{\kappa}_f$) is detected in the 
final state
\begin{equation}
 e (E_i, \vec{k}_i) + A \rightarrow e' (E_f, \vec{k}_f) + X,    \label{qe_e}
\end{equation}
and 
\begin{equation}
 \nu_l (\varepsilon_i, \vec{\kappa}_i) + A \rightarrow l^- (\varepsilon_f, \vec{\kappa}_f) + X,  \label{ccqe_nu}
\end{equation}
where $l$ represents $e$, $\mu$, or $\tau$. Further, $A$ is the nucleus in its ground state $|J_i, M_i\rangle$ 
and $X$ is the unobserved hadronic final state. 

The double differential cross section for electron and neutrino-nucleus scattering of Eqs.~(\ref{qe_e}) and ~(\ref{ccqe_nu})
can be expressed as
\begin {eqnarray}
&  & \left(\frac{d^2\sigma}{d{\omega}d{\Omega}}\right)_{e}~=~\frac{\alpha^{2}}{Q^{4}}~\left(\frac{2}{2J_i+1}\right)
~E_{f}k_{f}\cos^{2}(\theta/2) \nonumber \\                         
&  & ~~~~~~~~\times~\zeta^{2}\left(Z', E_{f}, q\right)~\left[\sum_{J=0}^{\infty}\sigma_{L,e}^{J}+
\sum_{J=1}^{\infty}\sigma_{T,e}^{J}\right], \label{ddiffcs_e} 
\end {eqnarray}
and
\begin {eqnarray}
&  & \left(\frac{d^2\sigma}{d{\omega}d{\Omega}}\right)_{\nu}~=~\frac{G_{F}^{2}~\cos^{2}{\theta_c}}{(4\pi)^2}~\left(\frac{2}{2J_i+1}\right)~
\varepsilon_{f}{{\kappa}_{f}} \nonumber \\
&  & ~~~~~~\times~\zeta^{2}\left(Z', \varepsilon_{f}, q\right)~\left[\sum_{J=0}^{\infty}\sigma_{CL,\nu}^{J}+
\sum_{J=1}^{\infty}\sigma_{T,\nu}^{J}\right], \label{ddiffcs_nu}
\end {eqnarray}
where $\alpha$ is the fine-structure constant, $G_{F}$ is the Fermi coupling constant, and $\theta_{c}$ is the Cabibbo angle. 
The direction of the outgoing lepton is described by the solid angle $\Omega$. The lepton-scattering angle is $\theta$, the transferred 
four-momentum is $q^{\mu}(\omega,\vec{q})$ and $Q^{2} = -q_{\mu}q^{\mu}$. Further, $\zeta(Z', E, q)$ 
is introduced in order to take into account the distortion of the lepton wave function in the Coulomb field generated by $Z'$ 
protons, within a modified effective momentum approximation~\cite{Engel:1998}.

\begin{figure}
\includegraphics[width=0.99\columnwidth]{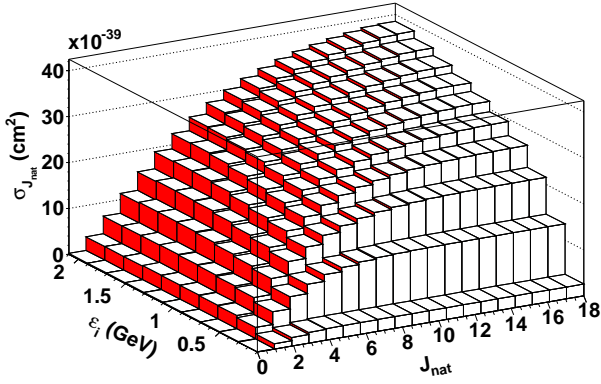}
\caption{(Color online) Multipole contributions (for natural parity transitions) to the 
cross section, as a function of the incoming neutrino energy. The $\sigma_{J_{nat}}$ denotes
the $^{12}$C$(\nu_{\mu},\mu^{-})$ cross section including all multipoles of the 
natural parity excitations up to $J_{nat}$.}
\label{fig_multipole}
\end{figure}

The $\sigma_{L,e}^J$ ($J$ denotes the multipole number) and $\sigma_{T,e}^J$ are the longitudinal and transverse components of the electron-nucleus
scattering
cross section, while $\sigma_{CL,\nu}^{J}$ and $\sigma_{T,\nu}^{J}$ are the Coulomb-longitudinal and transverse contributions 
of the neutrino-nucleus scattering cross section. In Fig.~\ref{fig_multipole}, we plot the 
strength obtained by adding the different multipole contributions to the cross section for incident neutrino energies from 0.1 to 2.0 GeV. 
Naturally, the higher 
the energy of the incident particle, the more multipoles contribute to the cross section. 
From the figure, one observes that for energies as low as 200 MeV, multipoles 
up to $J =$ 4 contribute. For energies as high as 2 GeV, multipoles up to $J =$16 need to 
be considered, and the relative weight of small $J$ contributions diminishes. 

The (Coulomb) longitudinal and transverse parts of the cross section both are composed of a kinematical factor $(v)$ and a response 
function $(R)$. The response function contains the full nuclear structure information. In the electron scattering case, the longitudinal 
($\sigma_{L,e}$) and transverse ($\sigma_{T,e}$) components of the cross section can be expressed as follows

\begin{eqnarray}
& & \sigma_{L,e} = v^{L}_e R^{L}_e, ~~~\sigma_{T,e} = v^{T}_e R^{T}_e, 
\end{eqnarray}
where the leptonic factors, $v^{L}_e$ and $v^{T}_e$, are given by
\begin{eqnarray}
& & v^{L}_e = \frac{Q^4}{|\vec{q}|^4}, ~~~v^{T}_e = \left[\frac{Q^2}{2|\vec{q}|^2}~+~\tan^2(\theta/2)\right].
\end{eqnarray}
Longitudinal $R^{L}_e$ and transverse $R^{T}_e$ response functions are defined as
\begin{eqnarray}
& & R^{L}_e = |\langle J_f|| \widehat{\mathcal{M}}_J^e(|\vec{q}|)|| J_i \rangle|^2, 
\end{eqnarray}
\begin{eqnarray}
& & R^{T}_e = \left[ |\langle J_f|| \widehat{\mathcal{J}}_J^{mag,e}(|\vec{q}|)|| J_i \rangle|^2 + 
|\langle J_f|| \widehat{\mathcal{J}}_J^{el,e}(|\vec{q}|)|| J_i \rangle|^2 \right]. \nonumber \\                  
\end{eqnarray}
Here $\widehat{\mathcal{M}}_J^e$, ${\widehat{\mathcal{J}}}_J^{mag,e}$ and ${\widehat{\mathcal{J}}}_J^{el,e}$ are the longitudinal, transverse 
magnetic and transverse electric operators, respectively~\cite{Connell:1972, Walecka:1995}. The $| J_i \rangle $ and 
$| J_f \rangle $ denote the initial and final state of the nucleus.

Similarly for neutrino-scattering processes, we express the Coulomb-longitudinal ($\sigma_{CL,\nu}$) and transverse
($\sigma_{T,\nu}$) parts of the cross section as follows: 
\begin{eqnarray}
\sigma_{CL,\nu} & = & \left[v^{\mathcal{M}}_\nu R^{\mathcal{M}}_\nu+ v^{\mathcal{L}}_\nu R^{\mathcal{L}}_\nu+
~2~ v^{\mathcal{M}\mathcal{L}}_\nu R^{\mathcal{M}\mathcal{L}}_\nu\right],  
\end{eqnarray}
\begin{eqnarray}
\sigma_{T,\nu}  & = & \left[v^{T}_\nu R^{T}_\nu+ 2 ~v^{TT}_\nu R^{TT}_\nu\right],   
\end{eqnarray}
where leptonic coefficients $v^{\mathcal{M}}_\nu$, $v^{\mathcal{L}}_\nu$, $v^{\mathcal{M\mathcal{L}}}_\nu$, $v^{T}_\nu$,
and $v^{TT}_\nu$ are given as
\begin{eqnarray}
 v^{\mathcal{M}}_\nu = \left[1+\frac{\kappa_f}{\varepsilon_f}\cos\theta\right], 
\end{eqnarray}
\begin{eqnarray}
 v^{\mathcal{L}}_\nu = \left[1+\frac{\kappa_f}{\varepsilon_f}\cos\theta- \frac{2 \varepsilon_i \varepsilon_{f}}
 {|\vec{q}|^2} {\left(\frac{\kappa_f}{\varepsilon_f}\right)}^2 \sin^{2}\theta\right], 
\end{eqnarray}
\begin{eqnarray}
 v^{\mathcal{M}\mathcal{L}}_\nu = \left[\frac{\omega}{|\vec{q}|}\left(1+\frac{\kappa_f}{\varepsilon_f}\cos\theta\right)+ 
 \frac{m_{l}^{2}}{\varepsilon_{f}|\vec{q}|}\right], 
\end{eqnarray}
\begin{eqnarray}
 v^{T}_\nu = \left[1-\frac{\kappa_f}{\varepsilon_f}\cos\theta+\frac{\varepsilon_i \varepsilon_{f}}{|\vec{q}|^2} 
 {\left(\frac{\kappa_f}{\varepsilon_f}\right)}^2 \sin^{2}\theta\right],  
\end{eqnarray}
 \begin{eqnarray}
 v^{TT}_\nu = \left[\frac{\varepsilon_i+\varepsilon_{f}}{|\vec{q}|}\left(1-\frac{\kappa_f}{\varepsilon_f}\cos\theta\right)-
 \frac{m_{l}^{2}}{\varepsilon_{f}|\vec{q}|}\right], 
\end{eqnarray}
and response functions $R^{\mathcal{M}}_\nu$,  $R^{\mathcal{L}}_\nu$,  $R^{\mathcal{ML}}_\nu$,  $R^{T}_\nu$, 
and $R^{TT}_\nu$ are defined as
\begin{eqnarray}
 R^{\mathcal{M}}_\nu = |\langle J_f|| \widehat{\mathcal{M}}_J^\nu(|\vec{q}|)|| J_i \rangle|^2, 
\end{eqnarray}
\begin{eqnarray}
 R^{\mathcal{L}}_\nu = |\langle J_f|| \widehat{\mathcal{L}}_J^\nu(|\vec{q}|)|| J_i \rangle|^2, 
\end{eqnarray}
\begin{eqnarray}
 R^{\mathcal{ML}}_\nu = ~\mathcal{R}\left[\langle J_f|| \widehat{\mathcal{L}}_J^\nu(|\vec{q}|)|| J_i \rangle 
 \langle J_f|| \widehat{\mathcal{M}}_J^\nu(|\vec{q}|)|| J_i \rangle^{\ast} \right],
\end{eqnarray}
\begin{eqnarray}
 R^{T}_\nu = \left[ |\langle J_f|| \widehat{\mathcal{J}}_J^{mag,\nu}(|\vec{q}|)|| J_i \rangle|^2 + 
 |\langle J_f|| \widehat{\mathcal{J}}_J^{el,\nu}(|\vec{q}|)|| J_i \rangle|^2 \right], \nonumber \\ 
\end{eqnarray}
\begin{eqnarray}
 R^{TT}_\nu = ~\mathcal{R}\left[\langle J_f|| \widehat{\mathcal{J}}_J^{mag,\nu}(|\vec{q}|)|| J_i \rangle 
 \langle J_f|| \widehat{\mathcal{J}}_J^{el,\nu}(|\vec{q}|)|| J_i \rangle^{\ast} \right]. \nonumber \\ 
\end{eqnarray}
Here $\widehat{\mathcal{M}}_J^\nu$, $\widehat{\mathcal{L}}_J^\nu$, $\widehat{\mathcal{J}}_J^{mag,\nu}$ and $\widehat{\mathcal{J}}_J^{el,\nu}$
are the Coulomb, longitudinal, transverse  magnetic, and transverse electric operators, respectively~\cite{Connell:1972, Walecka:1995}. 

\begin{figure}
\includegraphics[width=0.99\columnwidth]{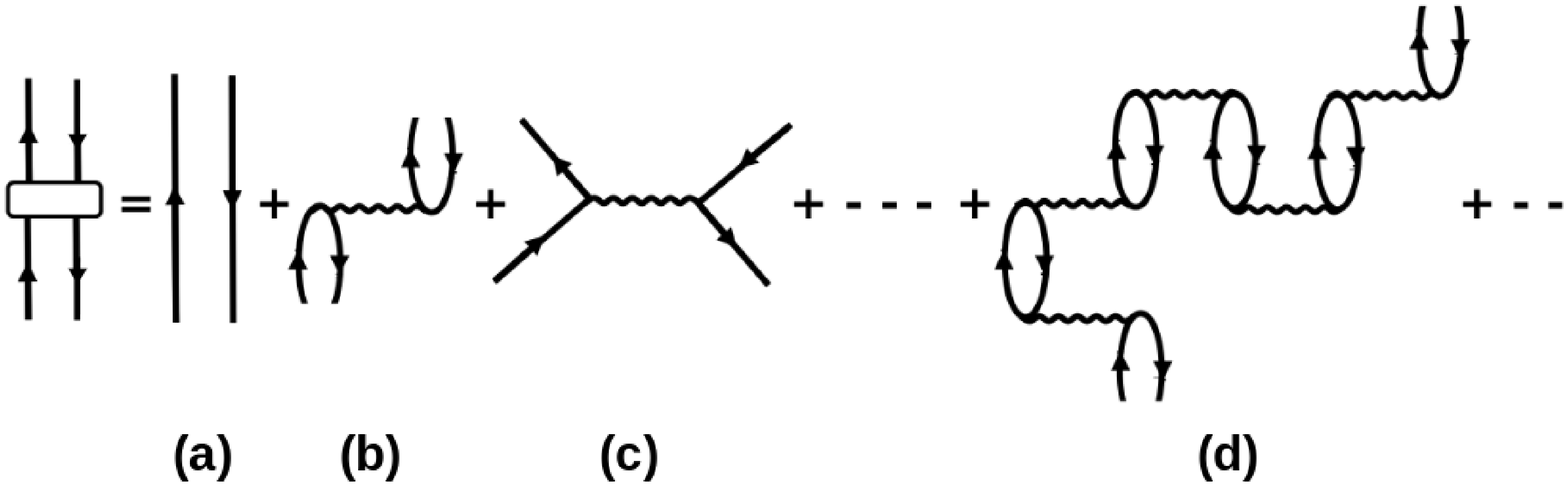}
\caption{Diagrammatic representation of the polarization propagator $\Pi^{(RPA)}$ for particle-hole states. Panel (a) corresponds to the 
unperturbed polarization propagator $\Pi^{(0)}$, (b) and (c) are the first-order direct and exchange RPA diagrams, 
and (d) represents a typical higher-order RPA diagram.}
\label{fig_propagator}
\end{figure}

To calculate the nuclear response functions, we use the CRPA approach which is described in detail in Refs.~\cite{Jan:1988, 
Jan:1989, Natalie:nc1999, Natalie:cc2002}. Here we will briefly present the essence of our model. We start by describing 
the nucleus within a mean-field (MF) approximation.
The MF potential is obtained by solving the Hartree-Fock (HF) equations 
with a Skyrme (SkE2) two-body interaction~\cite{Jan:1988,Jan:1989}. The sequential filling of the single-nucleon orbits 
automatically introduces Pauli-blocking. The continuum wave functions are obtained by integrating the 
positive-energy Schr\"odinger equation with appropriate boundary conditions.
In this manner, we account for the final-state interactions of the outgoing nucleon. 
Once
we have bound and continuum single-nucleon wave functions, we introduce the long-range correlations through a CRPA approach. 
We solve the CRPA equations with a Green's function formalism. The RPA describes a nuclear excited state as the linear combination of 
particle-hole ($ph^{-1}$) and hole-particle ($hp^{-1}$) excitations out of a correlated ground state
\begin{equation}
 \arrowvert \Psi_{RPA}^{C} \rangle = \sum_{C'} \left[ X_{C, C^{'}} ~ \arrowvert p'h'^{-1} \rangle - 
 ~Y_{C, C^{'}}~ \arrowvert h'p'^{-1} \rangle \right]~,
\end{equation}
where $C$ denotes the full set  of  quantum numbers representing an accessible channel. The Green's function approach 
allows one to treat the single-particle energy continuum exactly by treating the RPA equations in coordinate space. The RPA 
polarization propagator, obtained by the iteration of the first-order contributions to the particle-hole Green's function, is written as
\begin{eqnarray}
 & & \Pi^{(RPA)} (x_1,x_2;E_x) =  \Pi^{(0)} (x_1,x_2;E_x) \nonumber \\
&   &  + \frac{1}{\hbar} \int dx dx' \Pi^{0} (x_1,x;E_x)  
\tilde{V}(x, x') \Pi^{(RPA)} (x',x_2;E_x), \nonumber \\ && \label{propagator}
\end{eqnarray}
where $E_x$ is the excitation energy of the target nucleus and $x$ is a shorthand notation for the combination of the spatial, 
spin and isospin coordinates. The $\Pi^{(0)}$ in Eq.~(\ref{propagator}) corresponds 
to the HF contribution to the polarization propagator and $\tilde{V}$ denotes the antisymmetrized nucleon-nucleon 
interaction. The HF responses can be retrieved by switching off the second term in the above equation. Fig.~\ref{fig_propagator}
shows different components contributing to the polarization propagator.

\begin{figure}
\includegraphics[width=0.99\columnwidth]{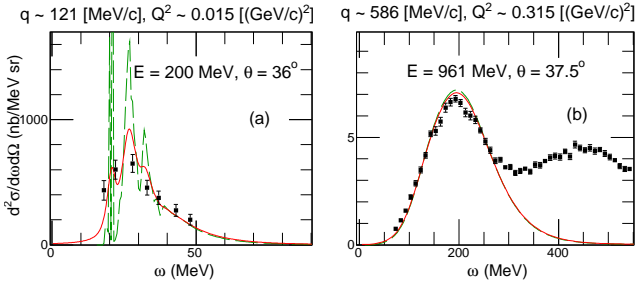}
\caption{(Color online) Comparison of $^{12}$C$(e,e')$ cross sections obtained with  
(full line) and without (dashed lines) the folding method. The experimental data are
from (a)~\cite{edata12C:Barreau} and (b)~\cite{edata12C:Sealock}.}
\label{fig_escattering_folding}
\end{figure}

A limitation of the RPA formalism is that the configuration space is restricted to 1p-1h excitations.  As a result
only the escape-width contribution to the final-state interaction is accounted for and the spreading width of the particle states is 
neglected. This affects the description of giant resonances in the CRPA formalism. The energy location of the giant resonance is generally 
well predicted
but the width is underestimated and the height of the response in the peak is overestimated. In order to remedy this, 
several methods have been proposed such as the folding procedure of Refs.~\cite{Smith:1988, DePace:1993, Co:2005, Amaro:2007}. 
Here, we use a simplified phenomenological approach where the modified response functions $R'(q,\omega')$ are obtained after folding 
the HF and CRPA response functions $R(q,\omega)$:
\begin{equation}
  R'(q,\omega') = \int_{-\infty}^{\infty} d\omega ~R(q,\omega)~ L(\omega, \omega'), \label{folding}
\end{equation}
with ($L$) a Lorentzian
\begin{equation}
 L(\omega, \omega') = \frac{1}{2\pi} \left[ \frac{\Gamma}{(\omega-\omega')^{2}+(\Gamma/2)^{2}} \right].
\end{equation}
We use an effective value of $\Gamma$ = 3 MeV which complies well with the predicted energy width in the giant-resonance region~[48], where
one expects the effect of the folding to be most important.
The overall effect of folding is a 
redistribution of strength
from peak to the tails. In line with the conclusions drawn in Refs.~\cite{Nieves:1997, Co:2005}, the energy integrated response functions  
are not much affected by the folding procedure of Eq.~(\ref{folding}). In Fig.~\ref{fig_escattering_folding},
we compare the ($e,e'$) cross sections obtained with and without folding. Figure~\ref{fig_escattering_folding} (a) clearly shows that in the giant-resonance region,  
the adopted folding procedure spreads the strength over a wider $\omega$ range, thereby considerably improving the quality of agreement with the 
data. At higher $\omega$ (Fig.~\ref{fig_escattering_folding} (b)) the effect of the folding is marginal. All computed cross-section results shown in the paper adopt the folding 
procedure of Eq.~(\ref{folding}).
\begin{figure*}
\begin{minipage}{\dimexpr\linewidth-0.50cm\relax}
\rotatebox{90}{\mbox{ {$d^{2}\sigma/d\omega d\Omega$~(nb/MeV sr)}}}
\vspace*{-21.7cm}\hspace*{16.7cm}
\end{minipage}%

\begin{minipage}{\dimexpr\linewidth-0.50cm\relax}
\includegraphics[width=0.90\textwidth]{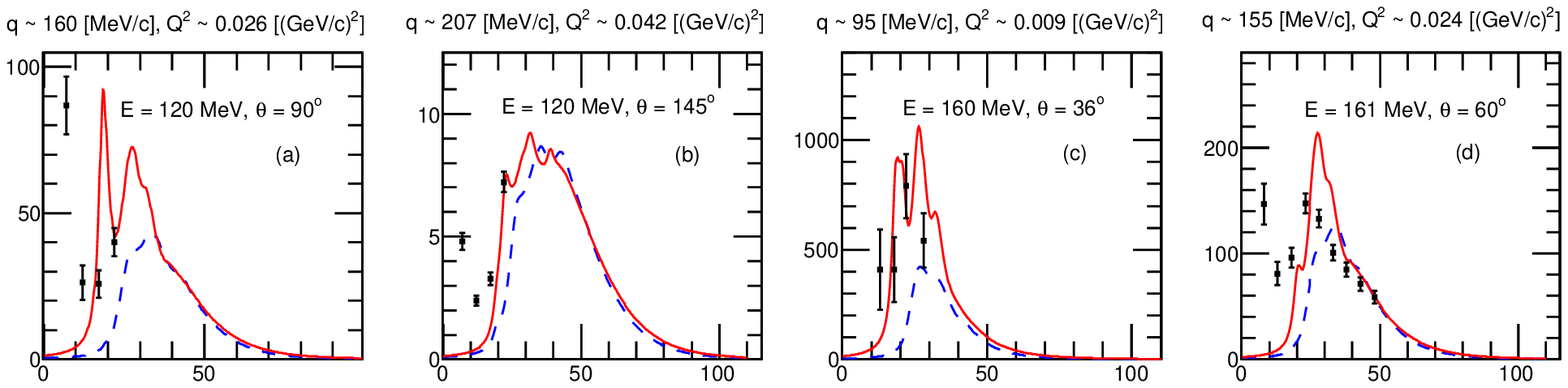}
\includegraphics[width=0.90\textwidth]{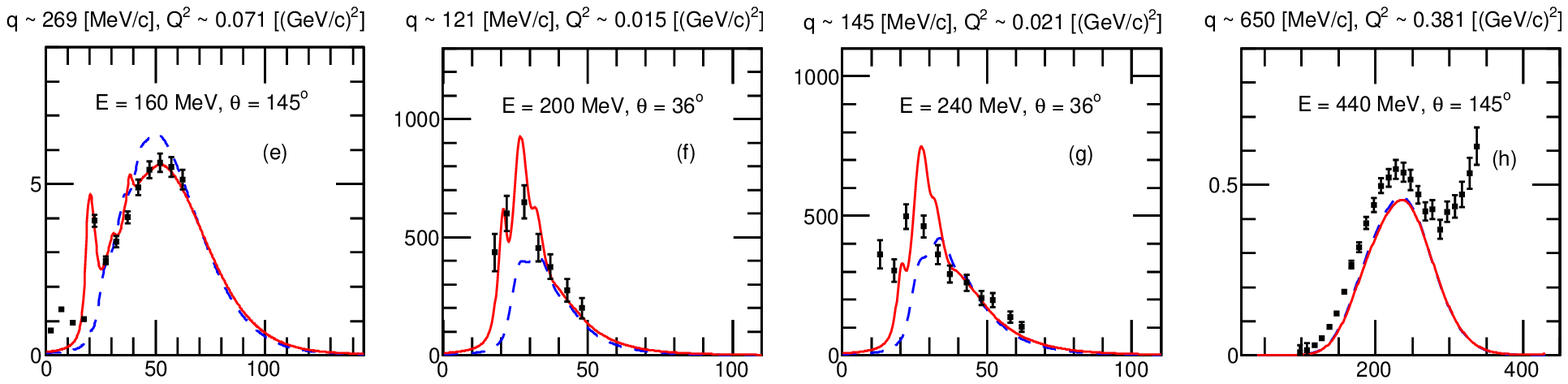}
\includegraphics[width=0.90\textwidth]{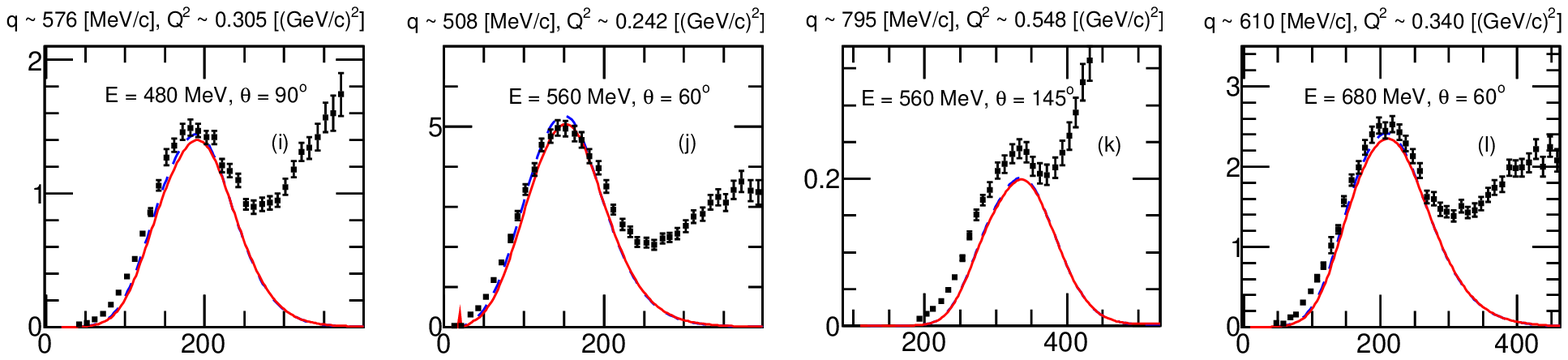}
\includegraphics[width=0.90\textwidth]{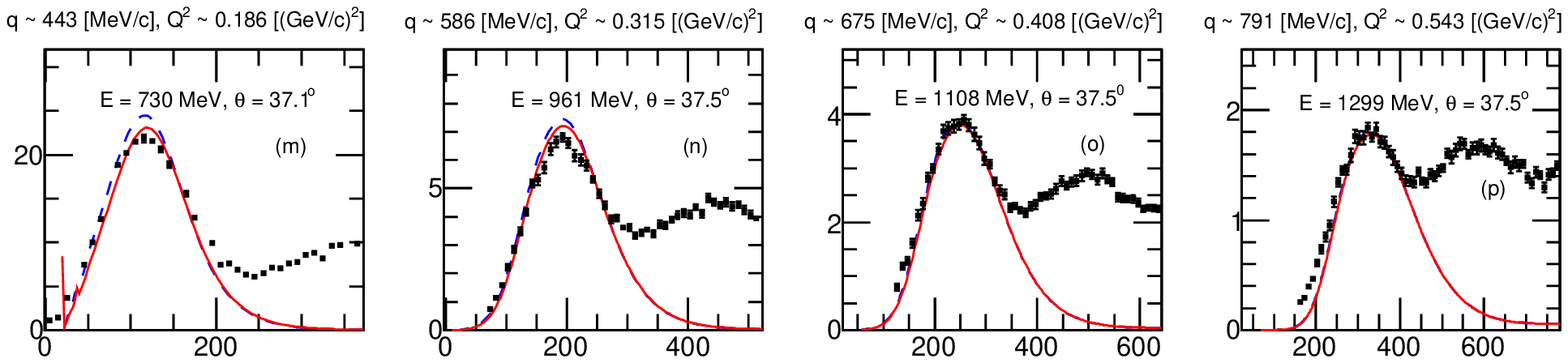}
\includegraphics[width=0.90\textwidth]{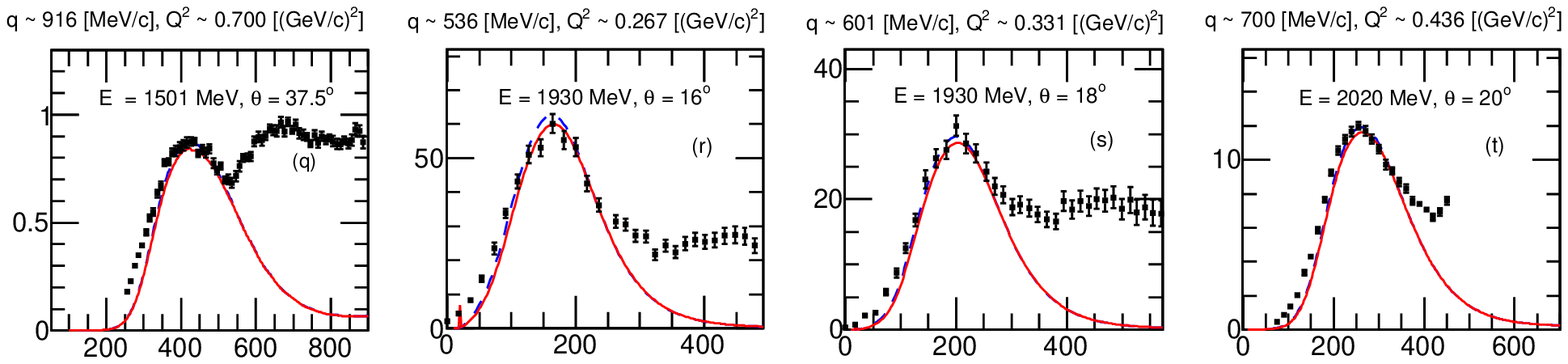}
\includegraphics[width=0.90\textwidth]{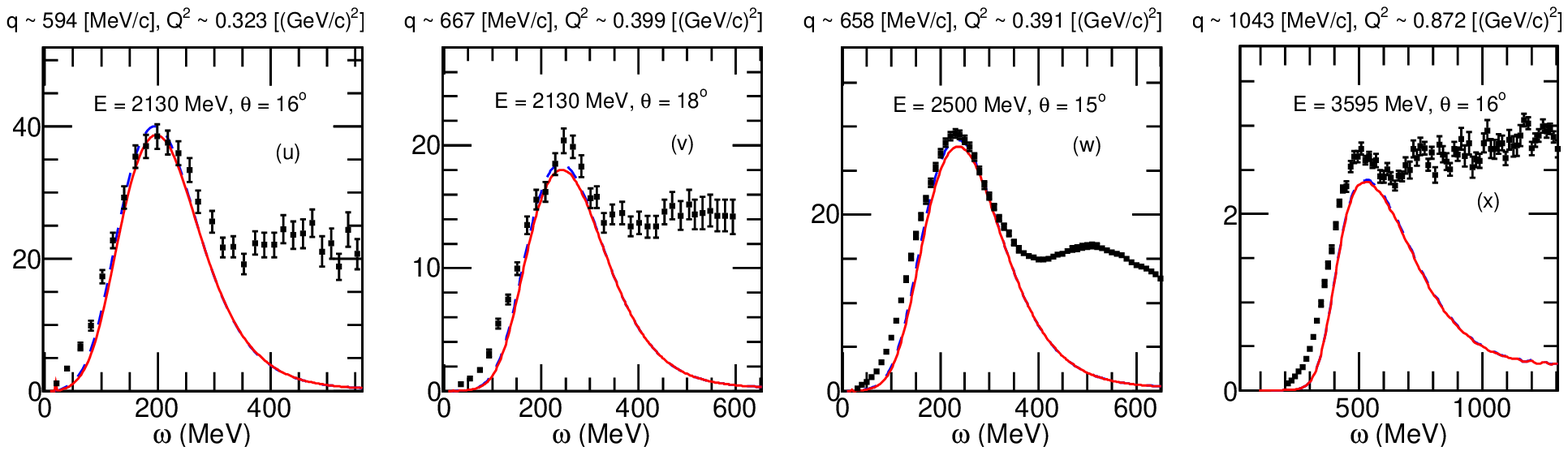}
\caption{(Color online) The double differential cross section for $^{12}$C($e,e'$). 
CRPA (solid lines) and  HF (dashed-lines) cross sections are compared with the data of 
Refs.~\cite{edata12C:Barreau, edata12C16O:O'Connell, edata12C:Sealock, edata12C:Bagdasaryan, 
edata12C:Day, edata12C:Zeller}. The $q$ and $Q^2$ values, on top of each panel, are calculated at 
quasielastic conditions $Q^2/(2M_{N} \omega) = 1$, with $M_{N}$ the nucleon mass.}
\label{fig_e_scattering_12C}
\end{minipage}
\end{figure*}

Our approach is self-consistent because we use the same SkE2 interaction in both the HF and CRPA equations. The parameters of the
momentum-dependent SkE2 force are optimized against ground-state and low-excitation energy properties~\cite{skyrme:physrep1987}.
Under those conditions the virtuality $Q^{2}$ of the nucleon-nucleon vertices is small. At high virtualities $Q^{2}$, the SkE2 force tends to be
unrealistically strong. We remedy this by introducing a dipole hadronic form factor at the nucleon-nucleon interaction vertices  
\begin{equation}
V(Q^2)~\rightarrow  V(Q^2=0)~\frac{1}{(1+\frac{Q^2}{\Lambda^2})^2}  \label{demp}
\end{equation}
where we introduced the free cut-off parameter $\Lambda$. 
We adopt $\Lambda~=$~455 MeV, a value which is optimized in a $\chi^2$ test of the comparison of $A(e,e')$ CRPA cross 
sections with the experimental data of Refs.~\cite{edata12C:Barreau, edata12C16O:O'Connell, edata12C:Sealock, edata12C:Bagdasaryan, 
edata12C:Day, edata12C:Zeller, edata16O:Anghinolfi, edata40Ca:Williamson}. In the $\chi^2$ test, we consider the theory-experiment comparison 
from low values of omega up to the maximum of the 
quasielastic peak. We have restricted our fit to the low-$\omega$ side of the quasielastic peak, because the high-$\omega$ side is subject to corrections 
stemming from intermediate $\Delta$ excitation, which is not included in our model.

The influence of the nuclear Coulomb field on the lepton is taken into account by means of an effective momentum approximation 
(EMA)~\cite{Engel:1998}. In order to take into account the reduced lepton wavelength, the three-momentum transfer is enhanced in an effective way
\begin{equation}
 q_{eff}~=~q+1.5~\left(\frac{Z'\alpha \hbar c}{R}\right),
\end{equation}
where $R = 1.24~A^{1/3}$~fm. The lepton wave functions are modified accordingly 
\begin{equation}
 \Psi_{l}^{eff}~=~\zeta(Z',E,q)~\Psi_{l}
\end{equation}
with
\begin{equation}
 \zeta(Z',E,q)~=~\sqrt{\frac{q_{eff}E_{eff}}{qE}},
\end{equation}
where $E$ ($E_{eff}$) is the energy (effective energy) of the outgoing lepton.

Our description of the nuclear dynamics is based on a nonrelativistic framework. For $q >$ 
500 MeV/c, the momentum of the emitted nucleon is comparable with its rest mass, and relativistic effects
become important. We have implemented relativistic corrections in an effective fashion, as suggested in 
Refs.~\cite{Donnelly:1998,Amaro:2005, Amaro:2007}. Those references show that a satisfactory description of relativistic 
effects can be achieved by following kinematic substitution in the nuclear response
\begin{equation}
  \lambda ~\rightarrow~ \lambda \left(1+\lambda\right),
\end{equation}
where $\lambda = \omega/2M_{N}$ and $M_N$ is the nucleon mass.  The above substitution produces a reduction of the width of the 
one-body responses and a shift in the peak towards smaller values of $\omega$.  The correction becomes sizable for $q$ $\gtrsim$ 500 MeV/c. 

\section {Results}\label{results}

To test our model, we start with the calculation of ($e,e'$) cross sections on 
different nuclei and the response functions for electron scattering off $^{12}$C, in subsection~\ref{results_electron}. We confront
our numerical results with the data of Refs.~\cite{edata12C:Barreau, edata12C16O:O'Connell, edata12C:Sealock, 
edata12C:Bagdasaryan, edata12C:Day, edata12C:Zeller, edata16O:Anghinolfi, edata40Ca:Williamson, edataRLRT12C:Jourdan}. 
We discuss the neutrino-scattering results in subsection~\ref{results_neutrino}.

\subsection{Electron scattering}\label{results_electron}

In this subsection, we present our results for the QE $A(e,e')$ cross sections. 
For any given $E_{i}$, the nuclear response depends on $q^{\mu}$. Energy transfers 
below the particle knockout threshold result in nuclear excitations in discrete states.
At slightly higher energies,
the giant dipole resonance (GDR) shows up. Only at substantially higher energy one can distinguish the peak corresponding to QE one-nucleon 
knockout. In an ideal case,
if an electron scatters from a free nucleon, one would expect a narrow peak at $\omega = Q^{2}/2M_{N}$. Deviations from that 
peak are due to the nuclear dynamics. The heavier the target nucleus, the wider the peak. 
The shift of the peak is due to nuclear binding and correlations.

\begin{figure}
\includegraphics[width=0.99\columnwidth]{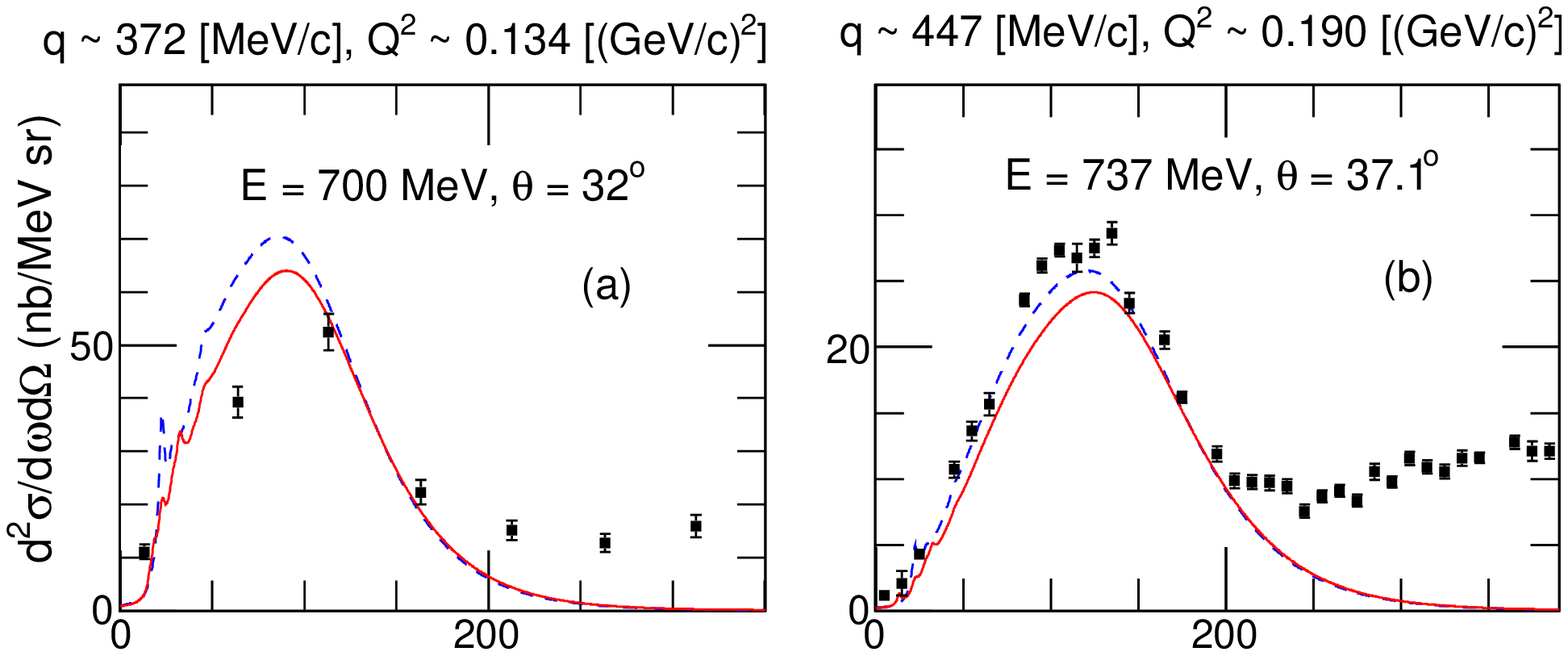}
\includegraphics[width=0.99\columnwidth]{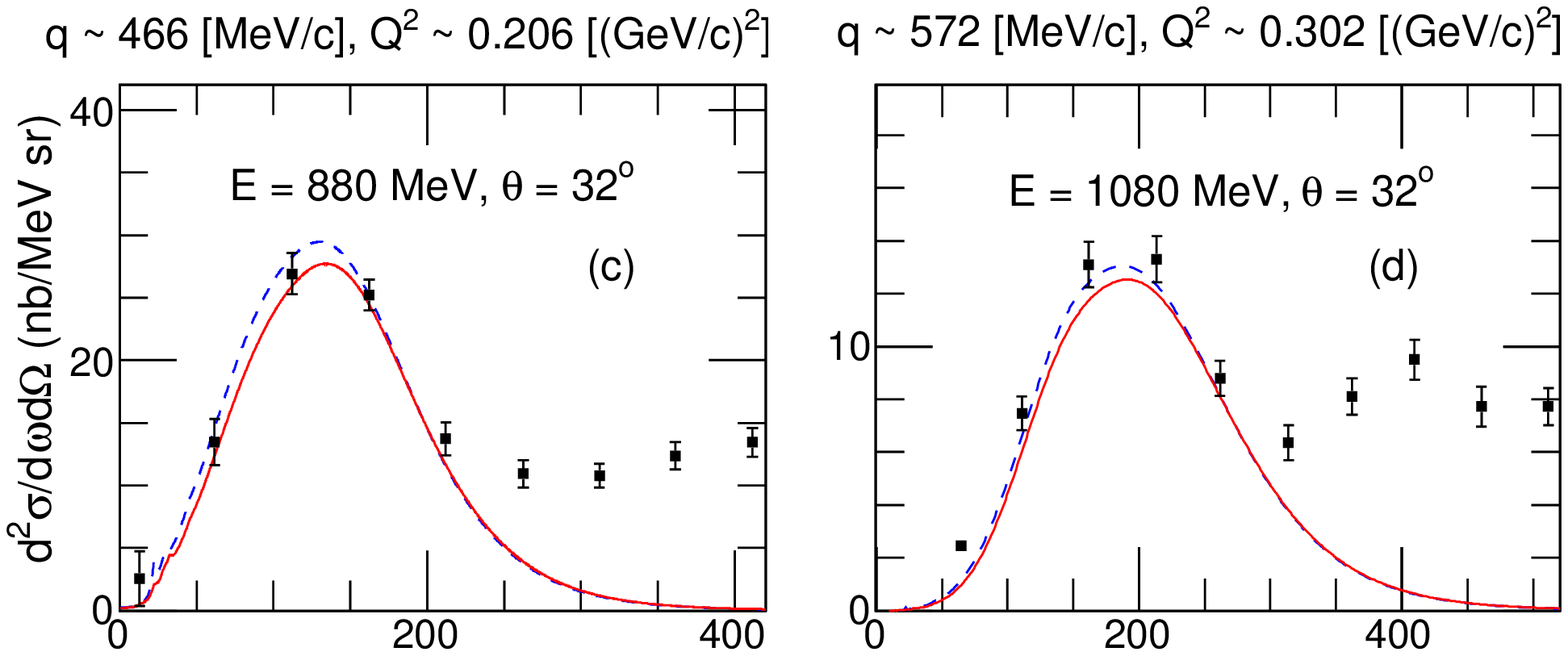}
\includegraphics[width=0.99\columnwidth]{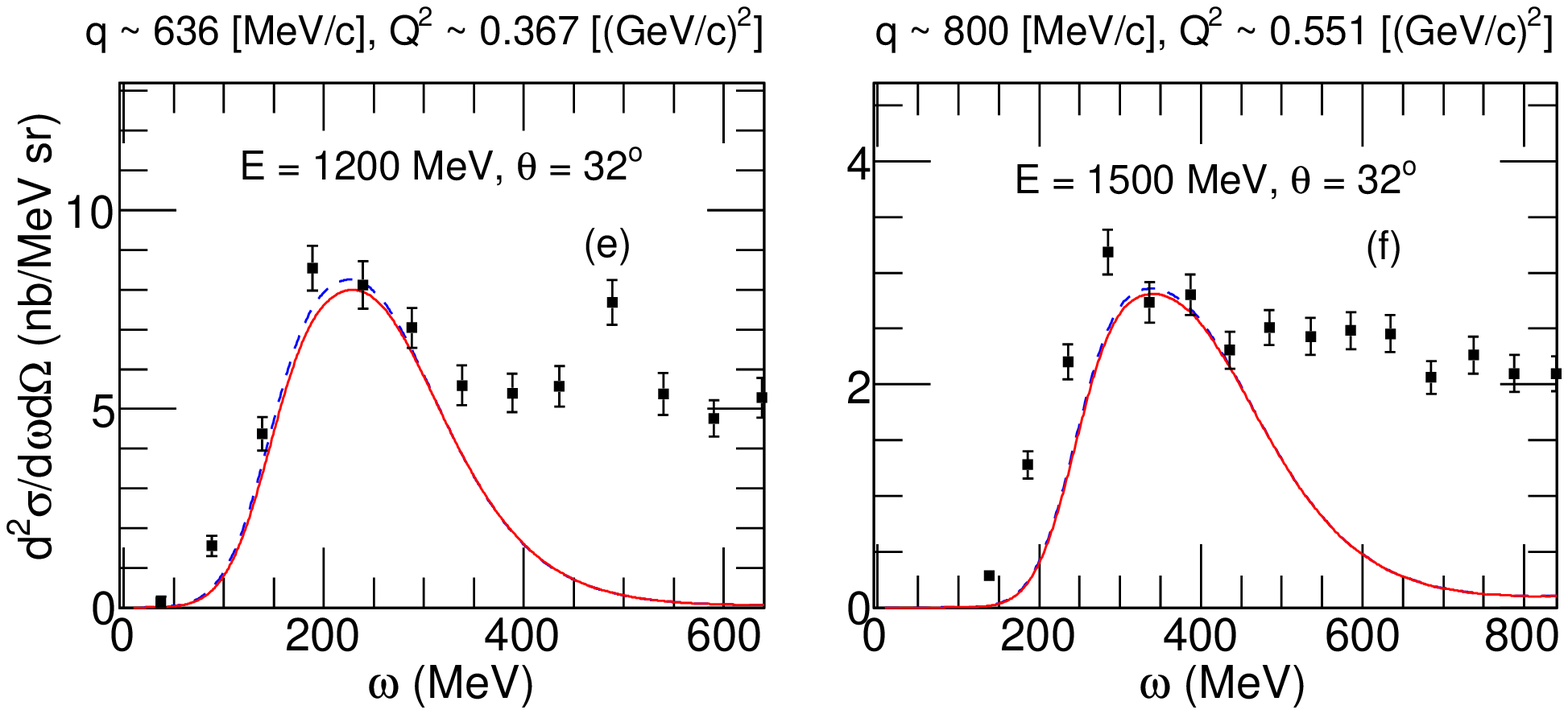}
\caption{(Color online) As in Fig.~\ref{fig_e_scattering_12C} but for $^{16}$O($e,e'$). The 
data are from Refs.~\cite{edata16O:Anghinolfi, edata12C16O:O'Connell}.}
\label{fig_e_scattering_16O}
\end{figure}
\begin{figure}
\includegraphics[width=0.99\columnwidth]{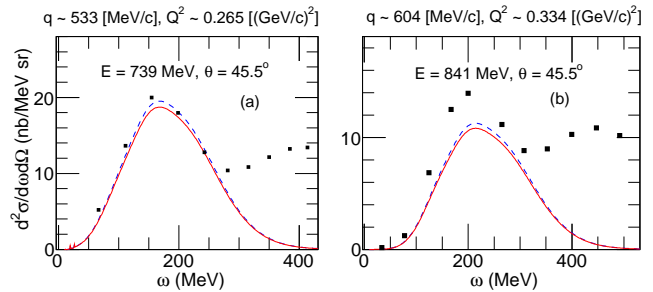}
\caption{(Color online) Same as Figs.~\ref{fig_e_scattering_12C} and \ref{fig_e_scattering_16O} 
but on a $^{40}$Ca target, the measurements are from Ref.~\cite{edata40Ca:Williamson}.} 
\label{fig_e_scattering_40Ca}
\end{figure}

\begin{figure}
\includegraphics[width=0.99\columnwidth]{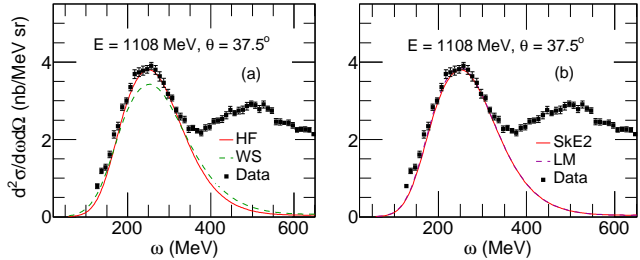}
\caption{(Color online) Comparison of cross sections obtained (on $^{12}$C) (a) with the HF and WS 
single-particle wave functions with SKE2 as residual interaction and (b) with the SKE2 and LM 
residual interaction with HF as single-particle wave functions.}
\label{fig_HF_WS_SKE2_LM}
\end{figure}

\begin{figure}
\includegraphics[width=0.99\columnwidth]{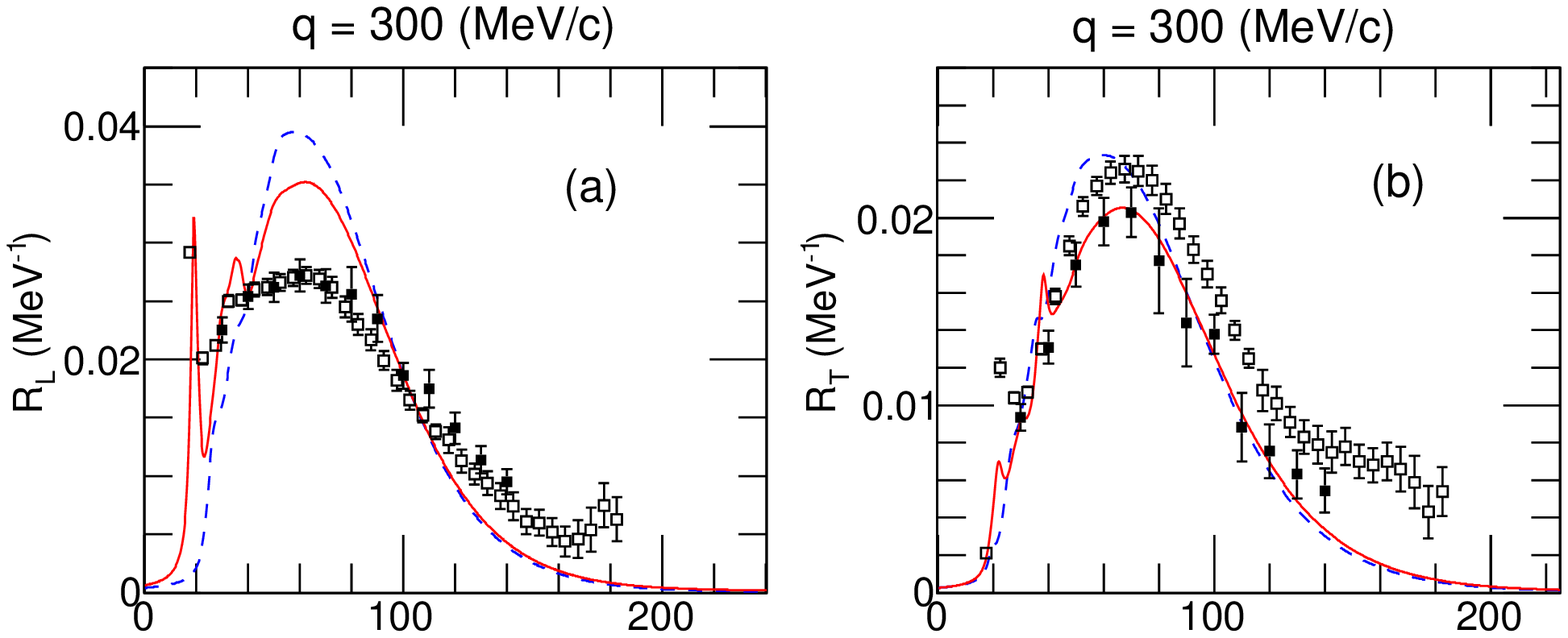}
\includegraphics[width=0.99\columnwidth]{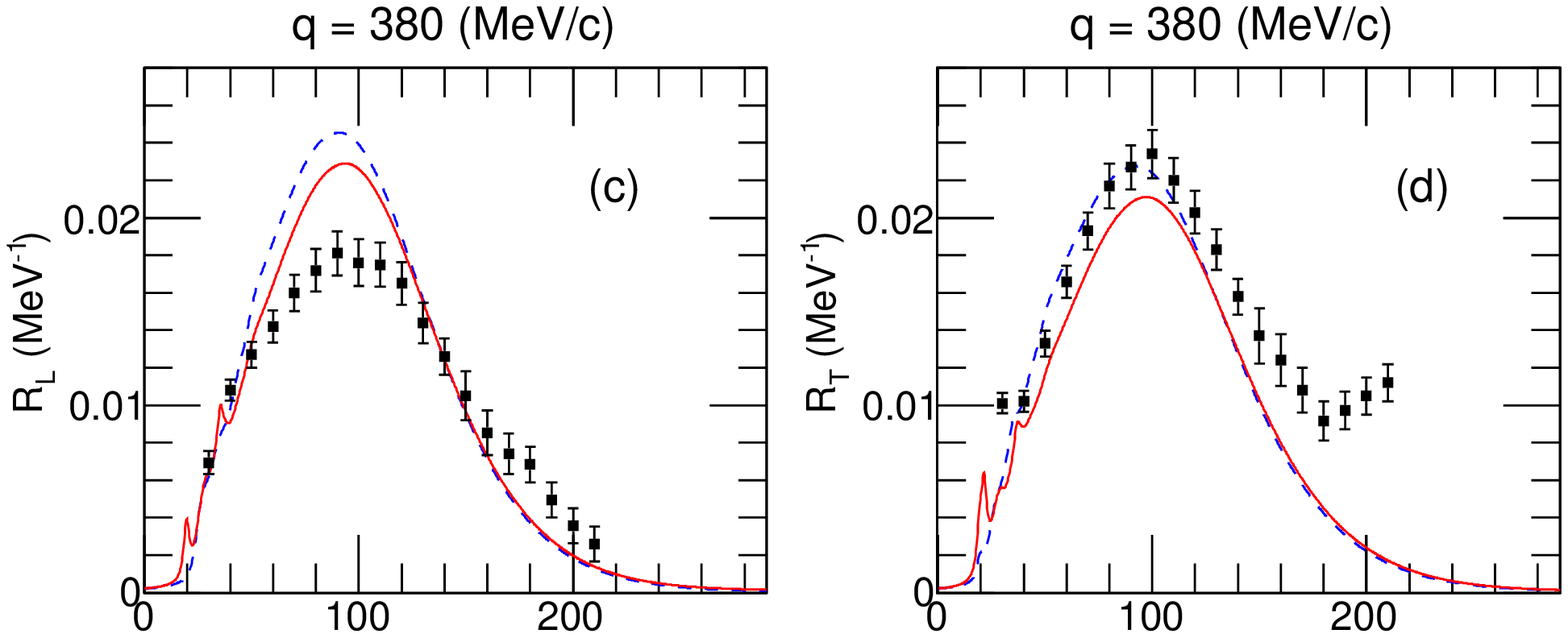}
\includegraphics[width=0.99\columnwidth]{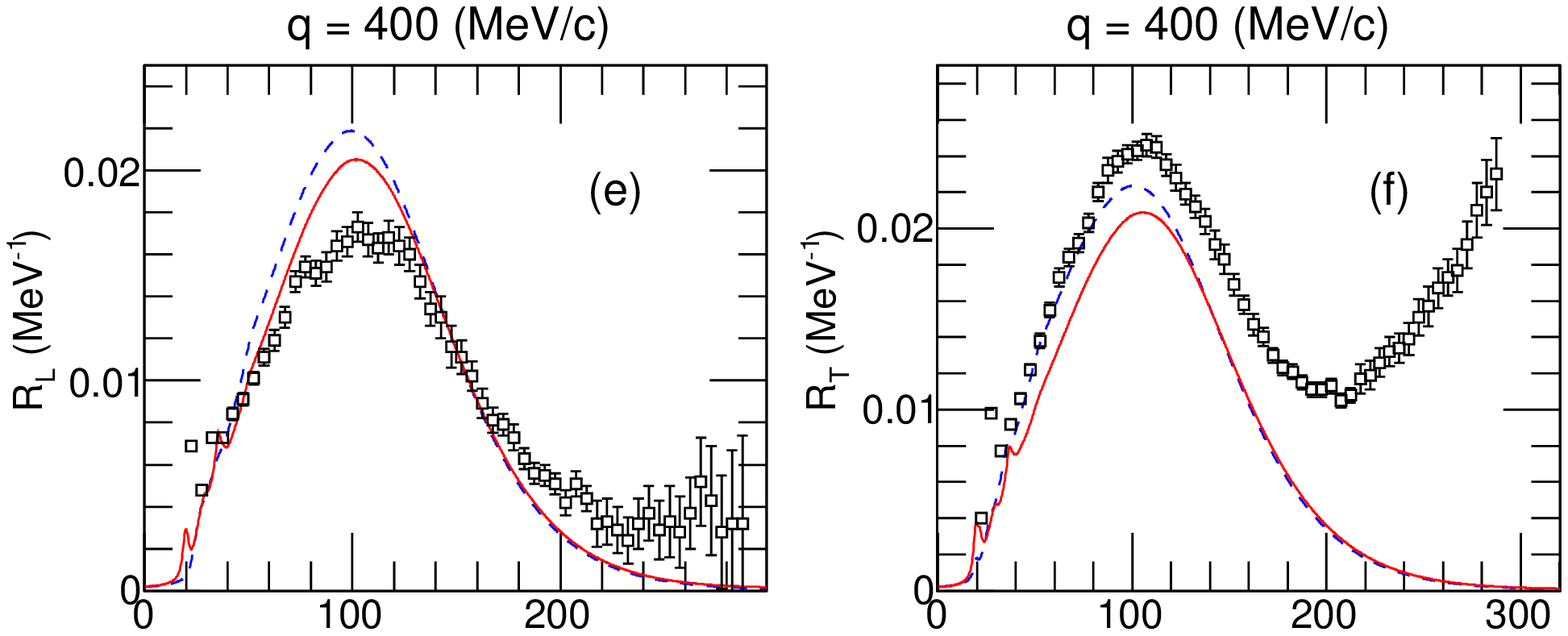}
\includegraphics[width=0.99\columnwidth]{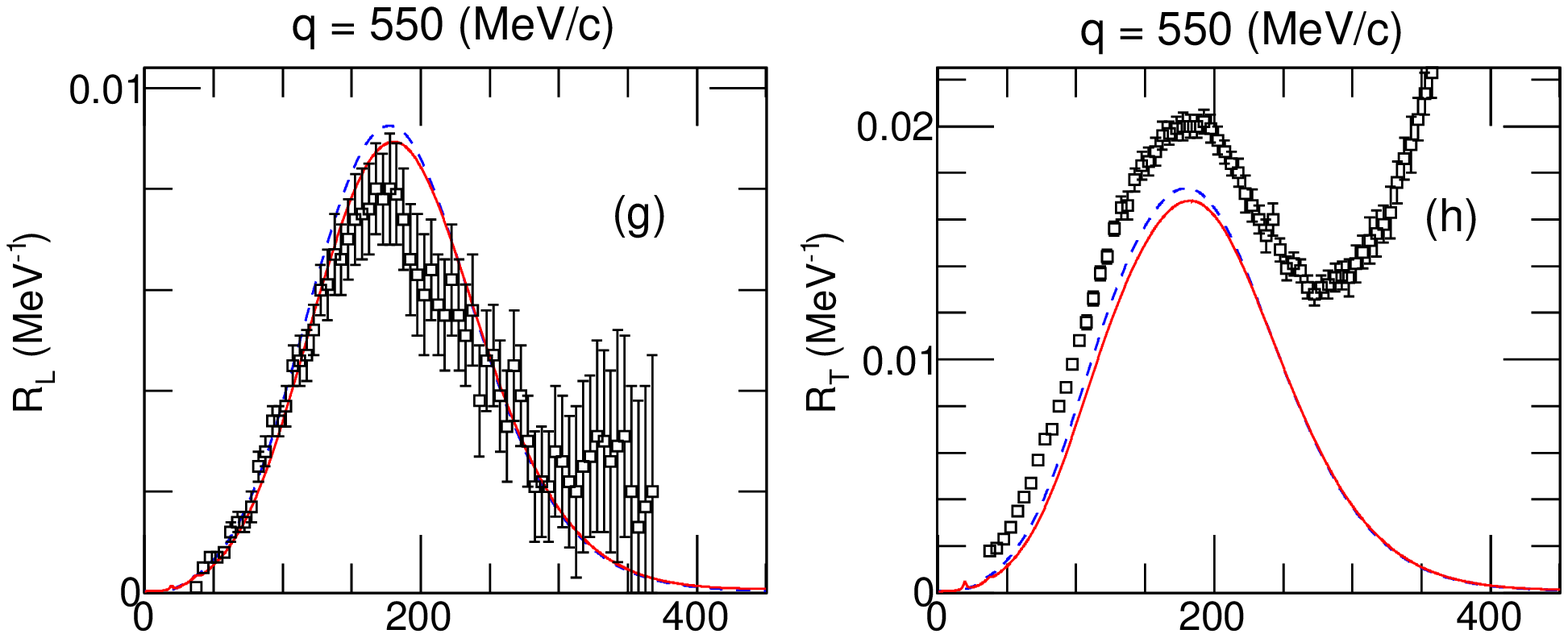}
\includegraphics[width=0.99\columnwidth]{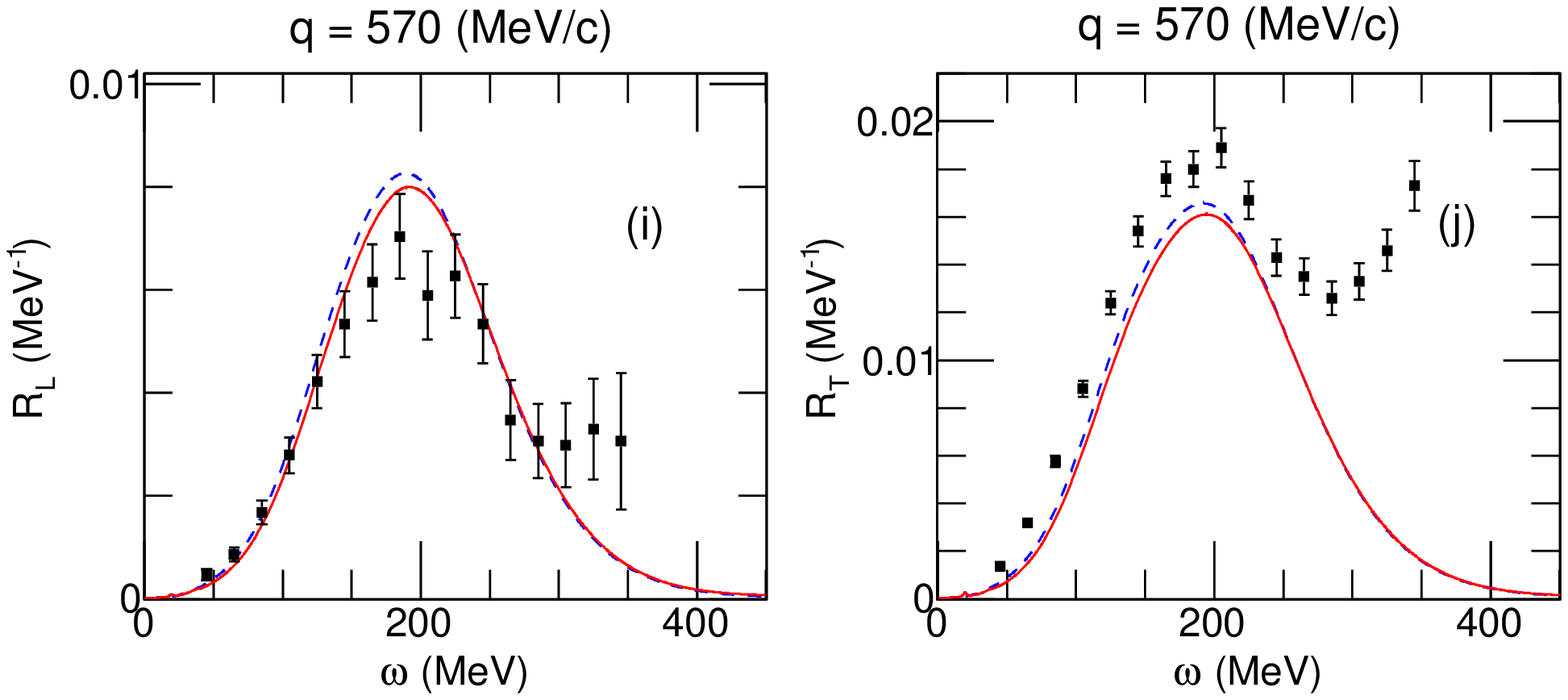}
\caption{(Color online) Longitudinal and transverse responses for
$^{12}$C($e,e'$), for different values of $q$. Solid lines are CRPA predictions and dashed lines are 
HF predictions. Experimental data are from Ref.~\cite{edataRLRT12C:Jourdan} (filled squares) and 
Ref.~\cite{edata12C:Barreau} (open squares).}
\label{fig_RL_RT}
\end{figure}

\begin{figure}
\includegraphics[width=0.99\columnwidth]{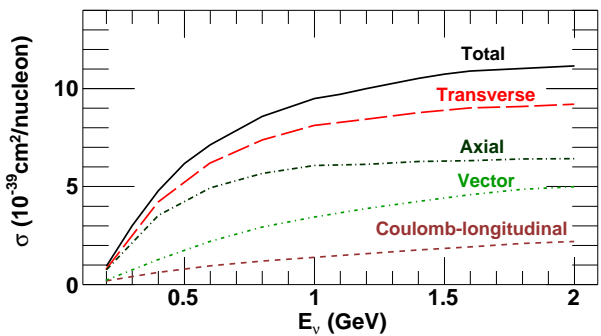}
\caption{(Color online) Different contributions to the total $^{12}$C($\nu_{\mu}, \mu^{-}$) 
cross section (per neutron) as a function of incoming neutrino energy. The sum of 
transverse and Coulomb-longitudinal (axial and vector) is the total cross section.}
\label{fig_cs_axial_vector_transverse_longitudinal}
\end{figure}
\begin{figure}
\includegraphics[width=0.99\columnwidth]{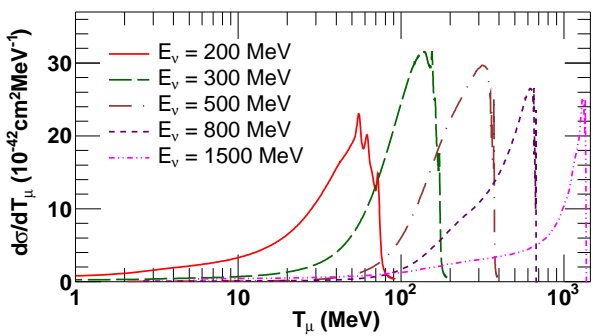}
\caption{(Color online) Cross section for $^{12}$C($\nu_{\mu}, \mu^{-}$) 
as a function of outgoing muon kinetic energy $T_{\mu}$, for different incoming neutrino energies.
Note the log scale on the horizontal axis. }
\label{fig_nu_dcs}
\end{figure}

\begin{figure*}
\includegraphics[width=0.325\textwidth]{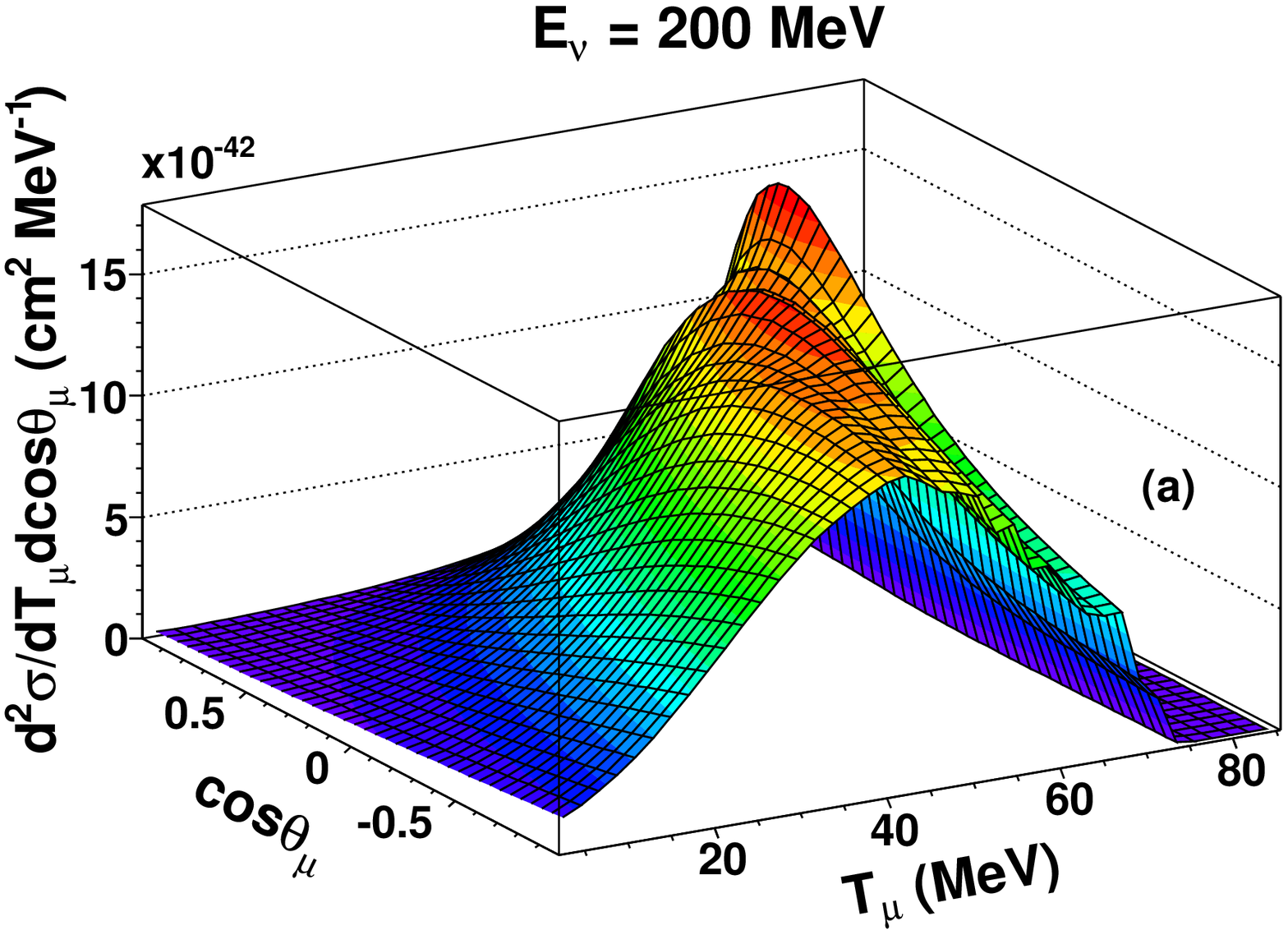}
\includegraphics[width=0.325\textwidth]{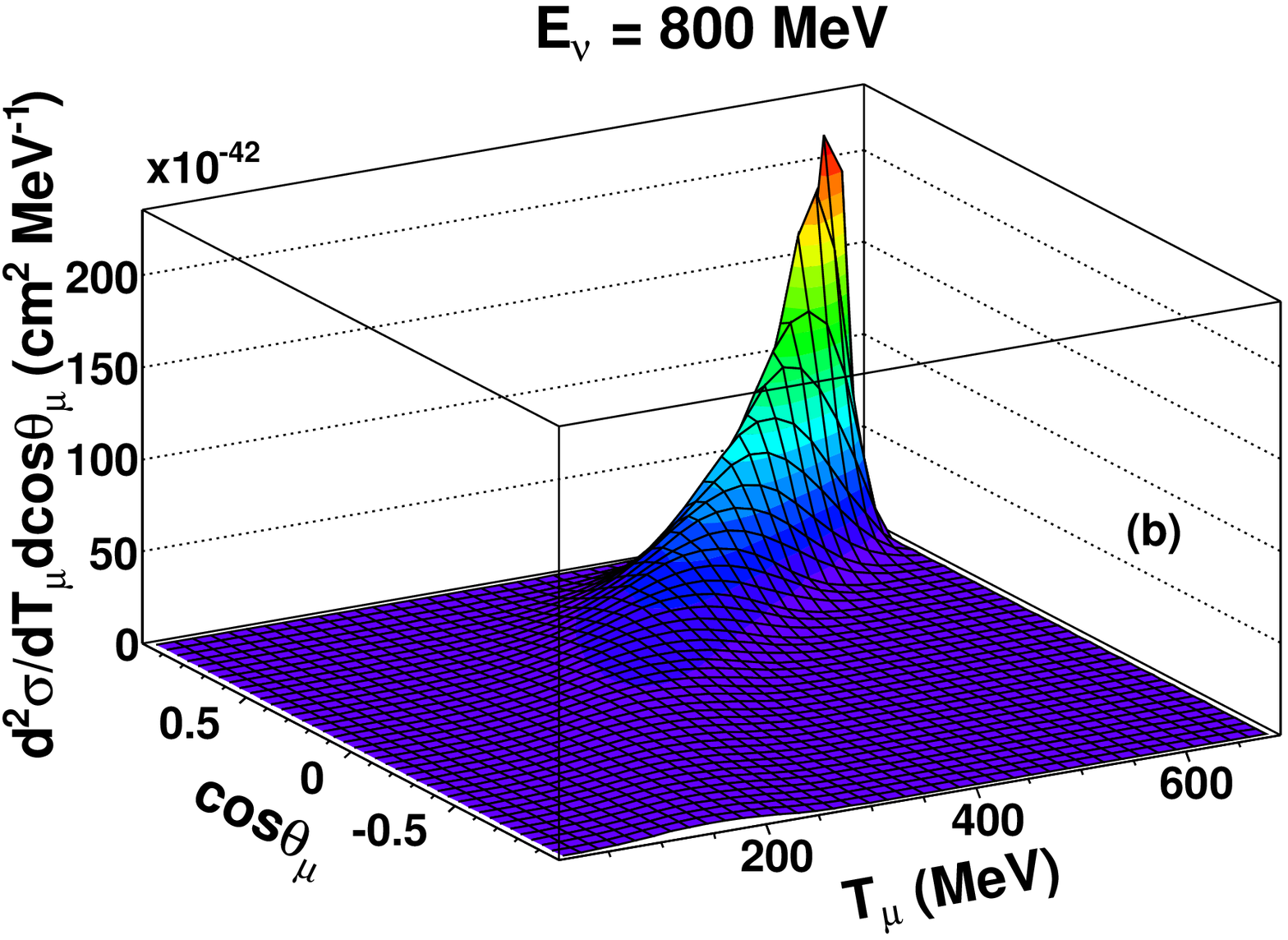}
\includegraphics[width=0.325\textwidth]{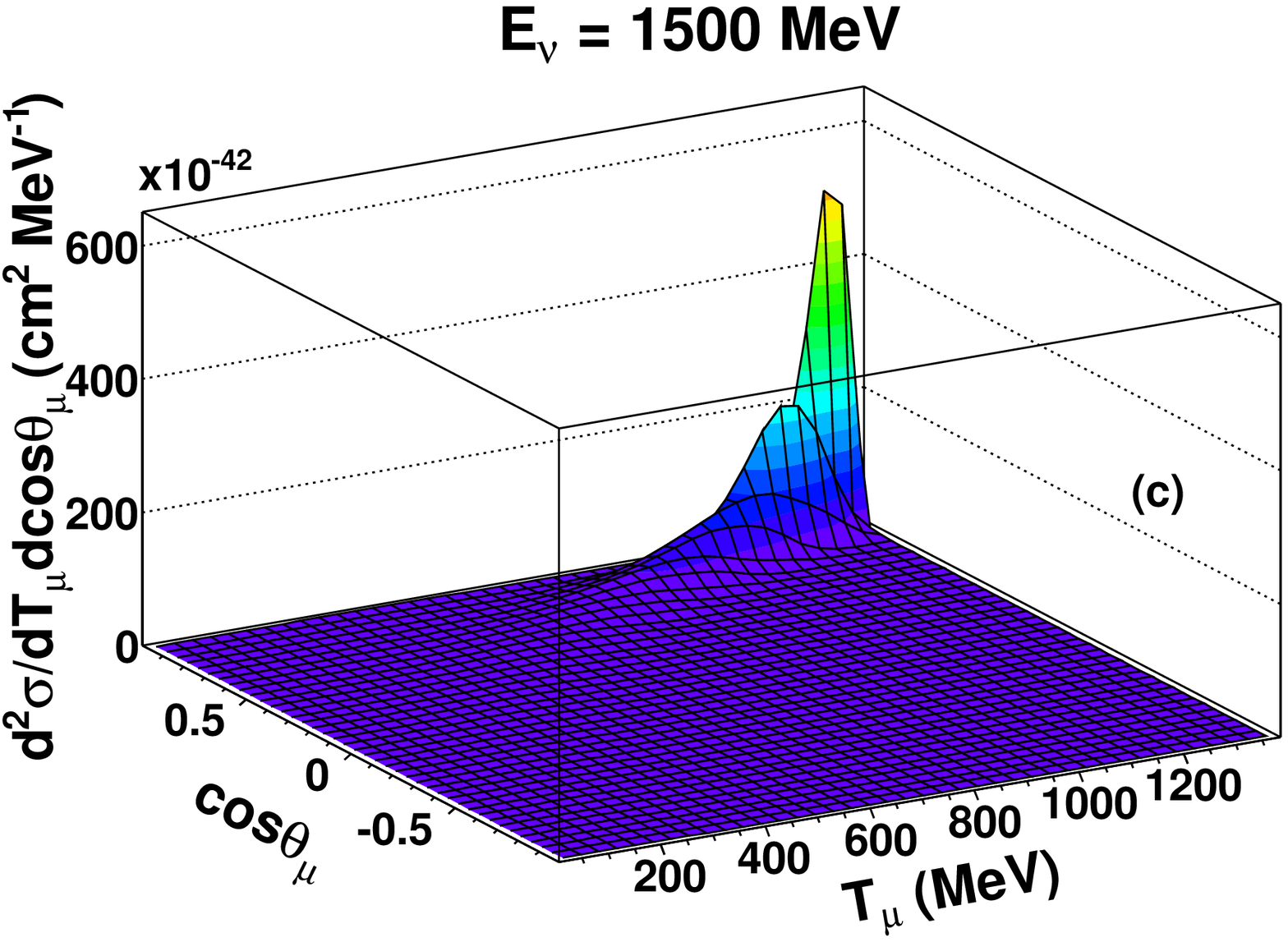}
\caption{(Color online) Double differential $^{12}$C($\nu_{\mu}, 
\mu^{-}$) cross sections plotted as a function of  $T_{\mu}$ and $\cos\theta_{\mu}$, for three neutrino 
energies.}
\label{fig_nu_ddcs}
\end{figure*}
\begin{figure}
\includegraphics[width=0.99\columnwidth]{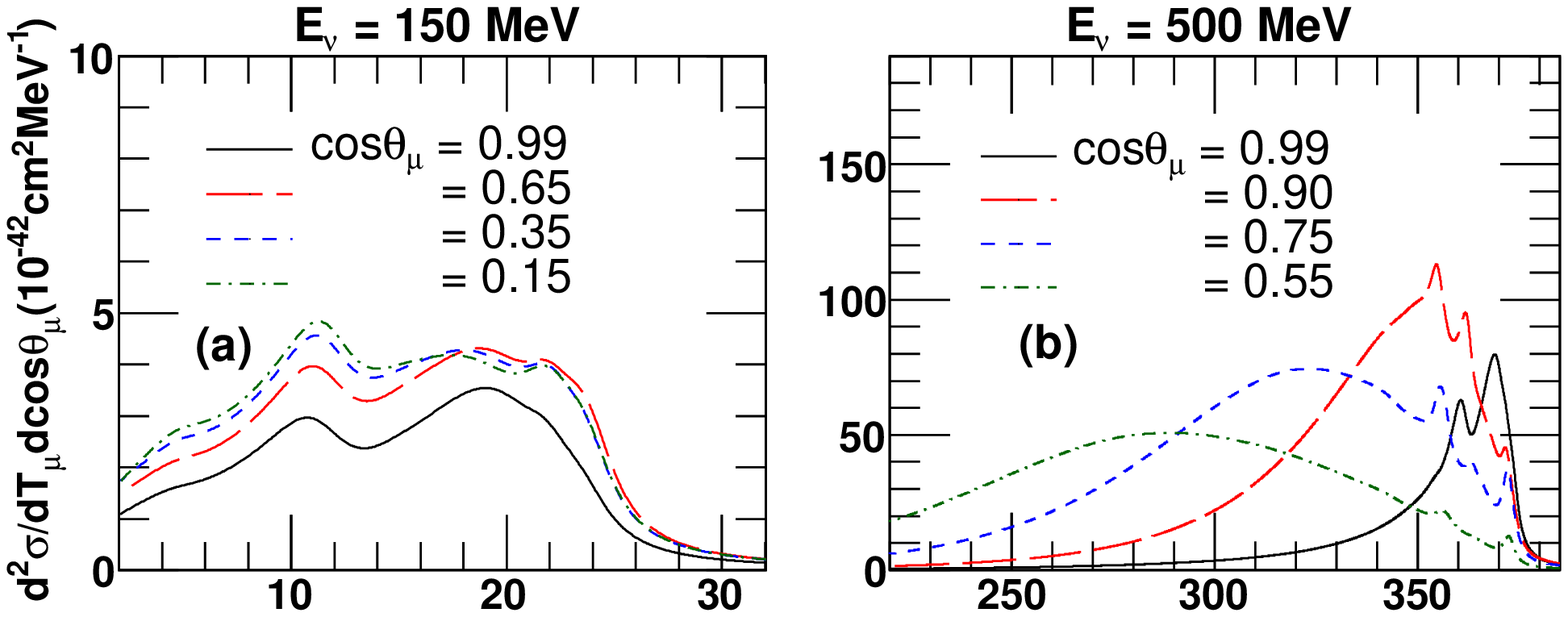}
\includegraphics[width=0.99\columnwidth]{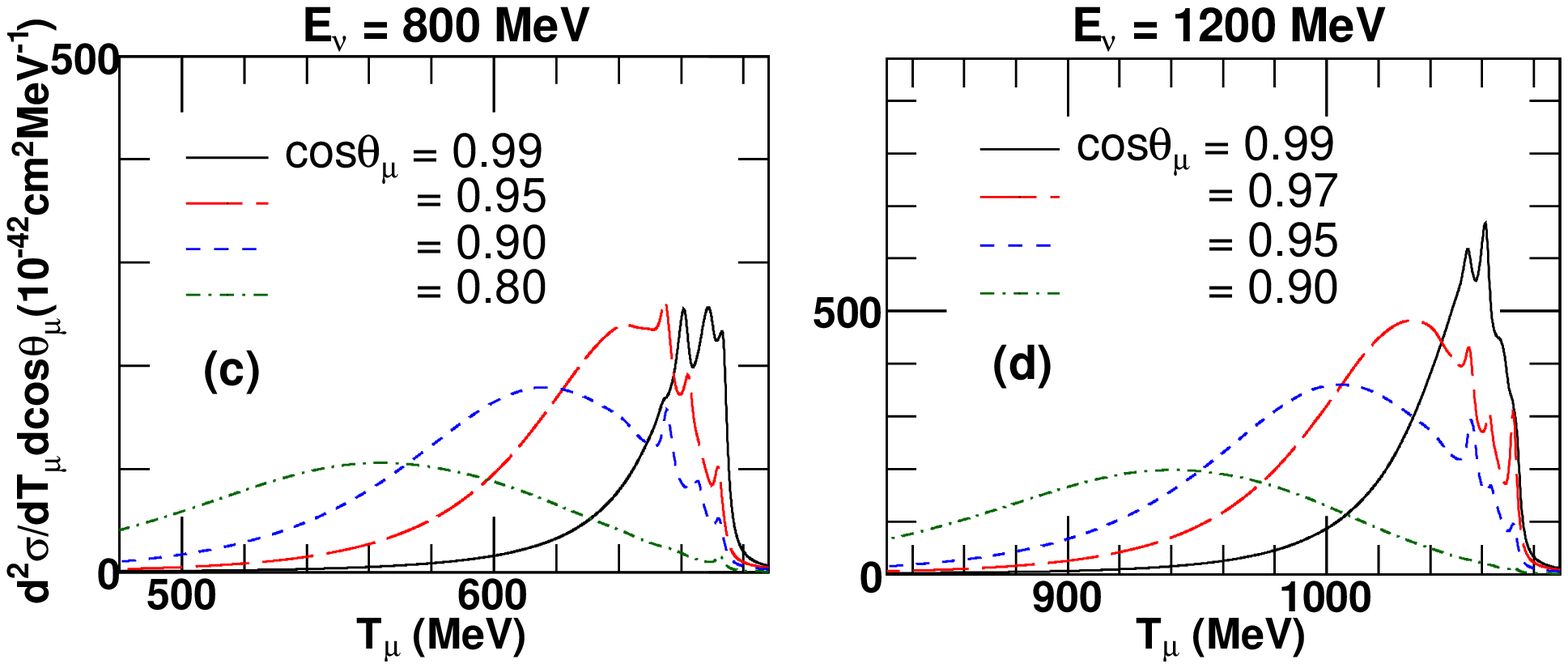}
\caption{(Color online) Low-energy excitations in double differential cross sections for
$^{12}$C($\nu_{\mu}, \mu^{-}$) plotted as a function of $T_{\mu}$, for different $\cos\theta_{\mu}$ 
values.}
\label{fig_nu_ddcs_lowenergy}
\end{figure}

\begin{figure}
\includegraphics[width=0.99\columnwidth]{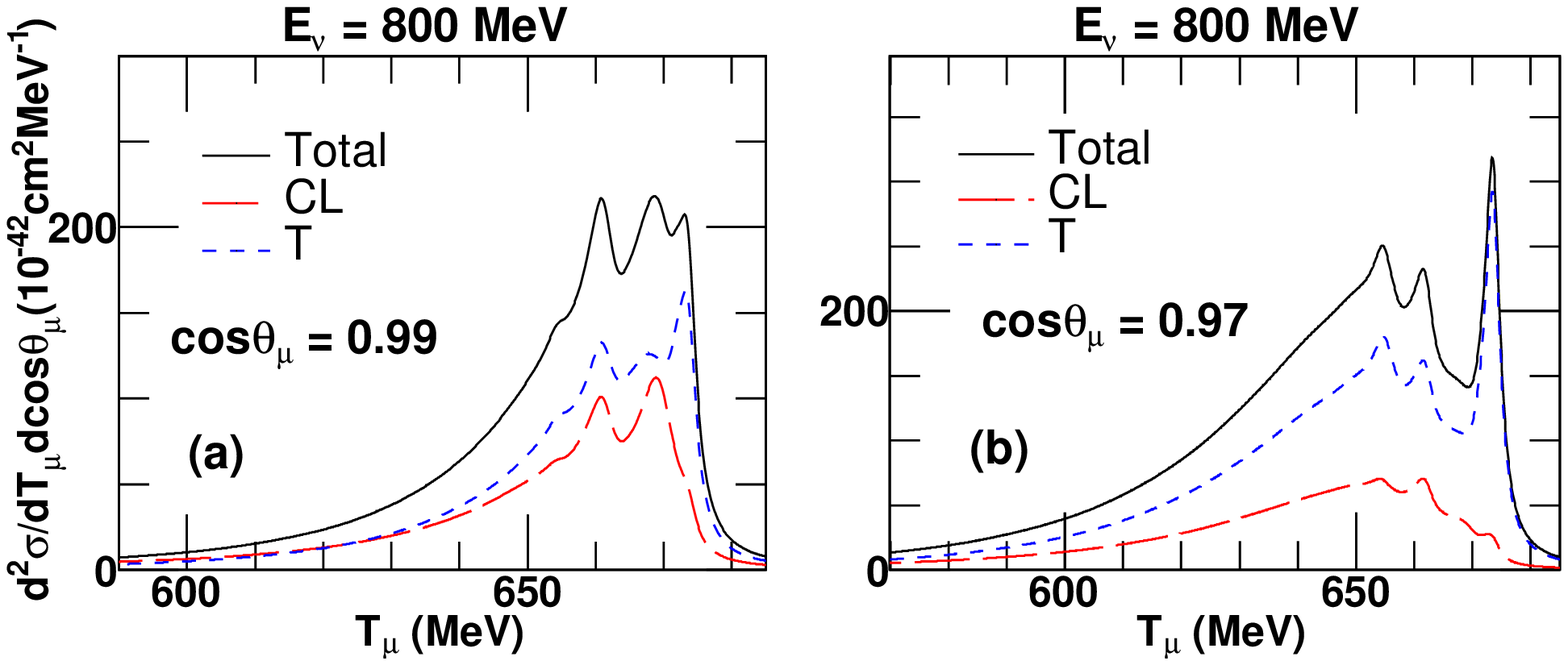}
\caption{(Color online) Coulomb-longitudinal (CL) and transverse (T) contributions to the double
differential cross sections, at $E_{\nu} = $ 800 MeV and two values of $\cos\theta_{\mu}$.}
\label{fig_nu_ddcs_CL_T}
\end{figure}

For the two vector form factors entering in the responses, we use the standard dipole parametrization of Ref.~\cite{dipoleformfactor:1991}. 
In Fig.~\ref{fig_e_scattering_12C} we present results of our numerical calculations for $^{12}$C($e,e'$).
We compare CRPA and HF predictions with the measurements performed at the Saclay Linear Accelerator~\cite{edata12C:Barreau},
Bates Linear Accelerator Center~\cite{edata12C16O:O'Connell}, Stanford Linear Accelerator Center~\cite{edata12C:Sealock, edata12C:Day},
Yerevan electron synchrotron~\cite{edata12C:Bagdasaryan}, and DESY~\cite{edata12C:Zeller}.
The comparison is performed over a 
broad range of three- and four-momentum transfers: 95 $\lesssim q \lesssim$ 1050 MeV/c, and
0.009 $\lesssim Q^{2} \lesssim$ 0.900 (GeV/c)$^{2}$. 
Our predictions are reasonably successful in describing the data over the broad kinematical range considered here.
Moreover, they compare favorably with the cross-section results of Refs.~\cite{Butkevich:2005, Ankowski:2014}.
The interesting feature 
of our results is the prediction of the nuclear excitations at small energy ($\omega <$~50 MeV) and 
momentum transfers ($q <$~300 MeV/c), well below the QE peak. This feature can be appreciated in 
Figs.~\ref{fig_e_scattering_12C} (a)-\ref{fig_e_scattering_12C} (g).
The HF and CRPA A($e,e'$) cross sections are identical for $Q^2 \gtrsim 0.25$ (GeV/c)$^{2}$.
The cross section drops by two orders of magnitude with the shift in scattering angle from 
$36^{\circ}$ to $145^{\circ}$, for a fixed energy, as evident from Figs.~\ref{fig_e_scattering_12C} (c)-\ref{fig_e_scattering_12C}(e) for 
an incoming energy of 160 MeV.
Even for higher incoming electron energies the cross-section measurements at smaller scattering angles are still dominated by QE processes. 
Obviously, the measured cross sections include contributions from channels beyond QE, like $\Delta$-excitations, evident 
as the second peak in the data, and 2p-2h contributions.  Our description is restricted to QE processes and further work is in 
progress on the role of processes beyond QE ones~\cite{tom}. 

The double differential $^{16}$O($e,e'$) cross sections are shown in Fig.~\ref{fig_e_scattering_16O}. 
Our numerical calculations reasonably describe the QE parts of the measurements performed at ADONE~\cite{edata16O:Anghinolfi} and 
at the Bates Linear Accelerator Center~\cite{edata12C16O:O'Connell}. Further, the calculations for the heavier target $^{40}$Ca 
are presented in Fig.~\ref{fig_e_scattering_40Ca}. Again, the comparison with the experimental data taken at Bates Linear Accelerator 
Center~\cite{edata40Ca:Williamson} is fair.

In Fig.~\ref{fig_HF_WS_SKE2_LM} we compare cross sections obtained with two different parametrizations of the 
single-nucleon wave functions and nucleon-nucleon residual interactions. The Landau Migdal (LM)~\cite{Co:1985} and 
SKE2~\cite{Jan:1989, skyrme:physrep1987}
yield similar cross sections while the use of the Woods-Saxon (WS)~\cite{Co:1984} wave function slightly shifts and reduces the 
strength of the cross section. This can be attributed
to the fact that the HF wave functions have larger high-momentum components than the WS ones.

The ($e,e'$) cross section receives contributions from the longitudinal and transverse components, as can be seen in
Eq.~(\ref{ddiffcs_e}).
A separation of these two response
functions provides further detail about the target dynamics.
It is worth mentioning that the experimental values of responses are 
extracted from a set of cross-section measurements using a Rosenbluth separation
\cite{RLRT:Rosenbluth1950}. The data of Ref.~\cite{edataRLRT12C:Jourdan} is determined by a reanalysis of the world data on ($e,e'$) 
cross sections. Interestingly, that resulted in a significant difference from the measurements of Ref.~\cite{edata12C:Barreau}, 
as can be seen in Fig.~\ref{fig_RL_RT} (b). 
The comparison between our predictions on $^{12}$C with the experimental data of 
Refs.~\cite{edata12C:Barreau, edataRLRT12C:Jourdan} is quite satisfactory. 
The longitudinal 
responses are overestimated and the transverse responses are usually 
underestimated. Our predictions are in line with those predicted in Ref.~\cite{edataRLRT12C:Jourdan} and with the
continuum shell model predictions of Ref.~\cite{Amaro:1999}.
It is long known, that the inclusion of processes involving meson exchange current (MEC) are needed to account for 
the transverse strength of the electromagnetic response~\cite{Donnelly:1978, Alberico:1984}. The calculations carried out on light nuclei 
overwhelmingly suggest that single-nucleon knockout processes, such as in this work, are dominant in the longitudinal channel 
while in the transverse channel two-nucleon processes provide substantial contributions. 

\subsection{Neutrino scattering}\label{results_neutrino}

The calculation of $^{12}$C($\nu_{l}, l^{-}$) response functions involve two vector form factors 
and one axial form factor. We use the BBBA05 parametrization of Ref.~\cite{BBBA05formfactor:2006} for the two vector form 
factors, and the standard dipole parametrization of the axial form factor with $M_{A} = $ 1.03
$\pm$ 0.02 GeV~\cite{Beringer:betadecay2012, Amsler:2008, Bernard:MaWorldAve2002}.

In Fig.~\ref{fig_cs_axial_vector_transverse_longitudinal} we display different contributions to the total $^{12}$C($\nu_{\mu}, \mu^{-}$) 
cross section, as a function of the incoming neutrino energy. The axial contribution is larger than the vector one. Related to this, neutrino
cross sections are dominated by the transverse current.

Electron-scattering cross-section measurements are typically
performed for a fixed incoming electron energy and scattering angle.  As neutrinos are produced as the secondary 
products of a decaying primary beam, the interacting neutrino's energy is not sharply defined.
The initial neutrino energy is reconstructed using the kinematics of the final outgoing lepton.   
This is a major source of uncertainty whereby nuclear structure can have an important influence.

The neutrino flux in oscillation experiments typically covers a wide energy range from about 100 MeV to a few GeV. 
The cross section measured at a single energy and scattering angle of the outgoing lepton picks up contributions from scattering 
processes at different energies, with varying weights. In Fig.~\ref{fig_nu_dcs}, we show the differential cross section 
(in outgoing muon energy) for 200 $ \lesssim E_{\nu} \lesssim $ 1500 MeV. It is evident from the figure that 
with increasing $E_{\nu}$ the strengths of the cross 
sections shift in muon energy. Also, there is a clear signature of the low-$\omega$ excitations even at neutrino energies around
the peak of the MiniBooNE and T2K $\nu_{\mu}$ spectra.

The measured cross sections are flux-folded double differential in outgoing muon kinetic energy $T_{\mu}$ and scattering angle $\cos\theta_{\mu}$.
To illustrate the low-energy excitations and general behavior of double differential $^{12}$C$(\nu_{\mu}, \mu^{-})$ cross sections at
fixed energies, we display in Fig.~\ref{fig_nu_ddcs} the double differential cross sections for
$E_{\nu} = $ 200, 800, and 1500 MeV. With the increase in incoming neutrino energy, the strength shifts in the forward direction and the width of 
giant resonances reduces.    
In Fig.~\ref{fig_nu_ddcs_lowenergy}, we plot the double differential cross section at different fixed values of $\cos\theta_{\mu}$.
For $E_{\nu} = $ 150 MeV, the double differential cross section is dominated by low-lying nuclear 
excitations, as evident from Fig.~\ref{fig_nu_ddcs_lowenergy} (a). For neutrino energies around the mean energy of the MiniBooNE~\cite{miniboone:ccqenu} 
and T2K~\cite{t2k:qenu} fluxes, $E_{\nu} = $ 800 MeV [Fig.~\ref{fig_nu_ddcs_lowenergy} (c)], the nuclear collective excitations are still sizable at forward muon 
scattering angles. The same feature is still visible for very forward scattering off neutrinos with an energy of 1200 MeV [Fig.~\ref{fig_nu_ddcs_lowenergy}(d)]. 
The contribution of collective excitations to neutrino-nucleus responses can not be accounted for within the RFG-based simulation codes.
As evident from the results presented here, they can account for non-negligible contributions to the signal even 
at higher neutrino energies. 

In Fig.~\ref{fig_nu_ddcs_CL_T}, we show the transverse and Coulomb-longitudinal contribution to the double differential cross
sections. For  
$\cos\theta_{\mu} = $ 0.99, the Coulomb-longitudinal contribution of the quasielastic cross section is comparable to the transverse one. 
The transverse contribution dominates the cross section as soon as one moves away from the very forward 
direction. This feature along with the giant-resonance contribution to forward-scattering 
cross sections accounts for most of the strength at very small momentum transfers. Theoretical models, which do not predict 
this behavior, tend to underestimate the cross
section for forward-scattering angles, as discussed in Ref.~\cite{Martini:2014}.


\section{Conclusions}\label{conclusions}

We presented a detailed discussion of CRPA predictions for quasielastic electron-nucleus 
and neutrino-nucleus responses. 

We assessed inclusive quasielastic electron-nucleus cross sections on $^{12}$C, 
$^{16}$O, and $^{40}$Ca. We consider momentum transfers over the broad range 95 $\lesssim q \lesssim$ 1050~MeV/c in combination with energy 
transfers which favor the quasielastic nucleon-knockout reaction process. We confronted our predictions with high-precision 
electron-scattering data. We separated the longitudinal and transverse responses on $^{12}$C, for 300 $\lesssim q \lesssim$ 
570 MeV/c, and compared them with the data. A reasonable overall description of the data, especially those corresponding with low-energy 
nuclear excitations, is reached. 

We calculated $^{12}$C($\nu_{\mu}, \mu^{-}$) cross sections, relevant for accelerator-based neutrino-oscillation 
experiments. We illustrated how low-energy nuclear excitations are induced by neutrinos. We paid special attention to 
contributions where nuclear-structure details become important, but remain unobserved in RFG-based models. 
We show that low-energy excitations can account for non-negligible contributions to the signal 
of accelerator-based neutrino-oscillation experiments, especially at forward neutrino-nucleus scattering. 


\acknowledgments
We thank Luis~Alvarez-Ruso and Teppei~Katori for useful discussions.
This research was funded by the Interuniversity Attraction Poles Programme initiated by the Belgian 
Science Policy Office, the Research Foundation Flanders (FWO-Flanders) and by the Erasmus Mundus External Cooperations Window's Eurindia Project.

\end{document}